\documentclass[fleqn,usenatbib]{mnras}

\usepackage{newtxtext,newtxmath}
\usepackage[T1]{fontenc}

\DeclareRobustCommand{\VAN}[3]{#2}
\let\VANthebibliography\thebibliography
\def\thebibliography{\DeclareRobustCommand{\VAN}[3]{##3}\VANthebibliography}

\usepackage{graphicx}	% Including figure files
\usepackage{amsmath}	% Advanced maths commands
\usepackage{url}
\usepackage[dvipsnames]{xcolor}
\usepackage{xcolor}

\newcommand\varcal[1]{\text{\usefont{OMS}{cmsy}{m}{n}#1}}

\title{Formation and evolution of boxy/peanut bulges in the Auriga cosmological simulations}

\author[P. D. López et al.]{
\hspace{-0.2cm} Paula D. López$^{1,2}$\thanks{E-mail: plopez@fcaglp.unlp.edu.ar},
Francesca Fragkoudi$^{2,3}$, %\thanks{E-mail: francesca.fragkoudi@durham.ac.uk},
Sofía A. Cora$^{1,4}$,
Cecilia Scannapieco$^{5,6}$,
Rüdiger Pakmor$^{7}$, \newauthor
Robert J. J. Grand$^{8}$,
Facundo Gómez$^{9}$
and
Federico Marinacci$^{10}$ 
\\
\\
$^{1}$Instituto de Astrof\'isica de La Plata (CCT La Plata, CONICET, UNLP), Observatorio Astron\'omico, Paseo del Bosque s/n, La Plata B1900FWA, Argentina\\
$^{2}$Institute for Computational Cosmology, Department of Physics, Durham University, South Road, Durham DH1 3LE, UK\\
$^{3}$School of Sciences, European University Cyprus, Diogenes Street, Engomi, 1516 Nicosia, Cyprus\\
$^{4}$Facultad de Ciencias Astron\'omicas y Geof\'isicas, Universidad Nacional de La Plata, Paseo del Bosque s/n, La Plata B1900FWA, Argentina\\
$^{5}$Universidad de Buenos Aires, Facultad de Ciencias Exactas y Naturales, Departamento de F\'isica, C1428EGA Buenos Aires, Argentina\\
$^{6}$Consejo Nacional de Investigaciones Cient\'ificas y Tecnol\'ogicas (CONICET), C1425FQB Buenos Aires, Argentina\\
$^{7}$Max-Planck-Institut für Astrophysik, Karl-Schwarzschild-Str. 1, 85748 Garching, Germany\\
$^{8}$Astrophysics Research Institute, Liverpool John Moores University, 146 Brownlow Hill, Liverpool, L3 5RF, UK\\
$^{9}$Departamento de Astronom\'ia, Universidad de La Serena, Av. Juan Cisternas 1200 Norte, La Serena, Chile\\
$^{10}$Department of Physics \& Astronomy “Augusto Righi”, University of Bologna, via Gobetti 93/2, 40129 Bologna, Italy 
}

\date{Accepted 2025 May 15. Received 2025 May 12; in original form 2025 March 20}

\pubyear{\the\year{}}

\begin{document}
\label{firstpage}
\pagerange{\pageref{firstpage}--\pageref{lastpage}}
\maketitle

\begin{abstract}
Boxy/peanut (b/p) or X-shaped bulges have been extensively explored with theory and numerical simulations of isolated galaxies. However, it is only recently that advances in hydrodynamical cosmological simulations have made it possible to explore b/p bulges in a cosmological setting, with much remaining to be understood about their formation and evolution. By using the Auriga magneto-hydrodynamical cosmological zoom-in simulations, we characterise the structural parameters of b/p bulges and how they form and evolve throughout cosmic history. 
We develop a method for estimating the b/p strength that allows us to identify the formation time and size of these structures. We find that b/p bulges in Auriga form between $\sim 1.1-1.6\, \rm Gyr$ after bar formation, following a `buckling' episode; some galaxies undergo multiple bucklings and events of b/p growth, with some b/p structures `dissolving' between buckling events. We find that at $z=0$, the b/p bulges have an extent of almost half the bar length.
Finally, we analyse the evolution of the b/p fraction over redshift, finding that at $z=0$, two thirds of galaxies host a bar, and of these, $45$ per cent have a b/p. This b/p fraction is within the observed range at $z=0$, although on the low end as compared to some observational studies. The b/p fraction decreases to $20$ per cent at $z=0.5$, and falls to zero at $z \sim 1$; this is in line with the observed trend of declining b/p fraction with redshift. We discuss possible culprits for the apparent mismatch in b/p occurrence between observations and cosmological simulations, what causes them to form (or not) in these simulations, and what this might reveal about models of galaxy formation and evolution.
\end{abstract}

\begin{keywords}
method: numerical -- galaxies: bars -- galaxies: bulge -- cosmology: theory
\end{keywords}

%%%%%%%%%%%%%%%%%%%%%%%%%%%%%%%%%%%%%%%%%%%%%%%%%%

%%%%%%%%%%%%%%%%% BODY OF PAPER %%%%%%%%%%%%%%%%%%

\section{Introduction}

Galaxies exhibit a variety of structural components, with central bulges being one of the most prominent ones. These structures are categorised into three types by \cite{Athanassoula_2005}: 
1) classical bulges, which are dispersion-dominated spheroids, composed mainly of old stars, 
2) nuclear discs (also commonly referred to as discy pseudobulges), which are flattened disc-like structures, and which often contain younger stellar populations, and
3) boxy/peanut (b/p) or X-shaped bulges. 
Early studies suggested that external torques \citep{May_1985} or galaxy mergers with specific configurations \citep{Binney_1985} could be the main formation mechanism of the latter. However, both these mechanisms require fine-tuning, while b/p bulges are common in edge-on galaxies (e.g. \citealt{Lutticke_2000}).
Several subsequent numerical studies have shown that these structures are in fact part of the bar seen edge-on and form spontaneously during the evolution of bars \citep{Combes_1981, Combes_1990, Pfenniger_1991, Raha_1991, Bureau_1999_1, Bureau_1999_2, Martinez-Valpuesta_2006, Ness_2012}.

Two primary formation mechanisms have been proposed as being responsible for forming b/p bulges (sometimes referred to as b/p’s, for simplicity): i) the so-called `buckling' of the bar, also known as the `fire hose' instability \citep{Toomre_1966, Raha_1991, Merritt_1994}, or ii) by the resonant excitation of the orbits of stars trapped within the bar \citep{Combes_1981, Combes_1990, Pfenniger_1991, Quillen_2014,Sellwood_2020}. 
Recently, \cite{Li_2023} proposed that the vertical buckling (or bending) of the bar is itself a resonant process.
Regarding the former, the bar becomes prone to vertical instabilities when the radial velocity dispersion exceeds the vertical velocity dispersion \citep{Araki_1985, Athanassoula_2007}. This causes the bar to bend, breaking its mid-plane symmetry and leading to a thickening of its inner regions (see for example \citealt{Cuomo_2023}). Using \textit{N}-body simulations, \cite{Martinez-Valpuesta_2006} showed that such buckling events can happen several times during the secular evolution of the bar. Meanwhile, resonance trapping refers to a gradual mechanism through which stars can be trapped in vertical resonances, leading to the formation of b/p bulges without the need for a buckling phase or a break in the mid-plane symmetry \citep{Quillen_2002}. 

There have been various methods proposed to test for the presence of a b/p bulge in numerical simulations and to characterise their features (e.g. \citealt{Debattista_2005}). \cite{Martinez-ValpuestaAthanassoula_2008} found correlations between the strength of bars and b/p bulges, indicating that stronger bars tend to have more pronounced b/p structures. They also observed a connection between bar strength and buckling episodes, with bars that experienced multiple buckling events being stronger. More recently, \cite{Ghosh_2024} used a suite of {\em N}-body models, featuring both thin and thick stellar discs, and found that in thicker discs the b/p bulges tend to be weaker and have a more extended length. 

Numerical studies of boxy/peanut bulges in galaxies have mostly involved isolated disc galaxies \citep{Combes_1990, Martinez-ValpuestaAthanassoula_2008, Fragkoudi_2017}, as these simulations provided the high resolution needed to examine their internal structures in detail. However, recent advances in cosmological simulations now make it possible to study the internal structures of disc galaxies within the full cosmological context of $\Lambda$-Cold Dark Matter (CDM).
This has led to significant advancements in the analysis of barred galaxies \citep{ScannapiecoAthanassoula_2012, Grand_2017, Algorry_2017, Spinoso_2017,Gargiulo_2019,Lokas_2020, Rosas-Guevara_2020, Fragkoudi_2021,Gargiulo_2022,Lopez_2024,Fragkoudi_2024}. In recent years, there have also been studies focused on the investigation of b/p structures in cosmological simulations. \cite{Fragkoudi_2020} used the Auriga simulations to analyse in detail the chemodynamic properties of five haloes with prominent bars and b/p structures. Through visual inspection, they found that the fraction of strong b/p's is around $30$ per cent at $z=0$ and discovered that young metal-rich populations produce prominent X-shaped structures, confirming previous results using isolated simulations \citep{Fragkoudi_2017, Debattista_2017, Athanassoula_2017}. Similarly, \cite{Blazquez-Calero_2020} used the Auriga simulations and identified six galaxies with b/p structures at $z=0$ through visual inspection of unsharp masked images, with two of these galaxies undergoing a buckling episode. They characterised these structures from an observational perspective, finding sizes comparable to those obtained from observations.
More recently, \cite{Anderson_2023} analysed barred galaxies with stellar mass $\rm log(M_{\star}/M_{\odot}) \geq 10.0$ at $z=0$ in the IllustrisTNG50 simulation \citep{TNGNelson2019, Pillepich_2019}. 
They distinguished those hosting a b/p structure from those that do not, 
and quantified the strength, frequency and sizes of these structures.

Observationally determining the fraction of b/p bulges (i.e. the number of barred galaxies that host a b/p structure) is somewhat complicated due to projection effects, since, on the one hand, it is challenging to ascertain whether an edge-on galaxy has a bar, while on the other, it is hard to determine whether a face-on galaxy possesses a b/p bulge. In the local Universe, it is observed that approximately two-thirds of disc galaxies possess a bar \citep{Eskridge_2000, Menendez-Delmestre_2007}, while there is a decrease in the bar fraction at higher redshifts \citep{Jogee_2004, Sheth_2008, Melvin_2014, Le_Conte_2024, Guo_2024}. Measurements of the b/p fraction in the local Universe have been conducted using different approaches. Some studies focus on edge-on galaxies, making use of the distinct boxy/peanut-shaped features that are more easily identifiable when the galaxy is viewed edge-on. \cite{Lutticke_2000} visually analysed edge-on galaxies in both optical and infrared bands from the Third Reference Catalogue of Bright Galaxies \citep[RC3,][]{de_Vaucouleurs_1991} finding a b/p fraction of $45$ per cent from a sample of $734$ galaxies, while \cite{Yoshino_Yamauchi_2015} used observations from the Sloan Digital Sky Survey (SDSS) Data Release 7 archive of 1716 edge-on galaxies, finding a b/p fraction of $22$ per cent in the {\em i}-band. 
Recently, \cite{Marchuk_2022} analysed images from the DESI Legacy Imaging Survey \citep{Dey_2019} creating a sample of $1925$ edge-on or near edge-on galaxies with b/p and X-shaped bulges, identified through residual images. They observed a significant increase in the frequency of b/p bulges in galaxies with masses above $\rm log(M_{\star}/M_{\odot}) \thickapprox 10.4$.

Other studies have concentrated on nearly face-on galaxies to assess the presence of b/p bulges, which are more challenging due to the orientation but provide complementary insights into the fraction of barred galaxies containing a b/p bulge. For example, \cite{Erwin_2017} examined a sample of $186$ disc galaxies from the RC3 catalogue and the Virgo Cluster Catalogue \citep{Binggeli_1985}. They found that $118$ galaxies of the sample were barred galaxies, and $84$ of these galaxies had a moderate inclination ($i \, \sim 40-70^{\circ}$) with the bar not oriented too close to the minor axis ($\Delta \rm PA < 60^{\circ}$), making them good for detecting possible b/p bulges. Their method involves studying the isophotes of galaxies, taking into account that the projection of the b/p bulge appears thick, resulting in a box-shaped isophote, while the outer part of the bar has a thinner projection, which they refer to as "spurs". Based on this observation, galaxies that meet the inclination criteria and exhibit isophotes with a box+spurs feature are classified as having b/p bulges. Their study revealed a strong dependence on stellar mass: only $12$ per cent of barred galaxies with masses less than $\rm log(M_{\star}/M_{\odot}) \thickapprox 10.4$ have b/p bulges, whereas about $80$ per cent of barred galaxies with masses greater than this threshold have b/p bulges. 
\cite{Kruk_2019} analysed a sample of galaxies from the Hubble Space Telescope COSMOS survey and SDSS using the same approach as the one developed by \cite{Erwin_2017}. They investigated the evolution of the b/p bulge fraction in barred galaxies with redshift, finding a $31$ per cent fraction at $z=0$ that decreases to $\sim 0$ at $z=1$. The authors argue that these detections represent lower limits for the b/p fraction, as they included galaxies with orientations that are not favourable for detecting b/p bulges, thereby underestimating the observed quantities. By applying a correction to their results, they find a b/p fraction of $\sim 69$ per cent at $z=0$, which drops to less than $\sim 10$ per cent at $z=1$. While the exact fraction of b/p's in the local Universe and across cosmic history is still to be fully quantified, it is clear that these structures are commonly found in spiral galaxies in the Universe.

Here we are interested in studying the formation and evolution of b/p bulges in a cosmological context, using the Auriga cosmological simulations. Our goal is to investigate how and when they form, as well as characterising their evolution over time.
The paper is organised as follows. In Section \ref{Sec:Simulation&Sample}, we describe the simulations used, present the global properties of the galaxies in our sample, and describe the method used to identify the b/p bulges. In Section \ref{Sec:bp_formation}, we study the formation time of b/p's and the fraction of galaxies hosting them across redshifts. In Section \ref{Sec:bar_bp_evolution}, we examine the time evolution of the sizes of the b/p bulges and the relation with the sizes of the bars. In Section \ref{Sec:buckling_episodes}, we study the buckling episodes experienced by the bars. In Section \ref{Sec:Discussion}, we discuss our results. Finally, in Section \ref{Sec:Summ&Conclu}, we present the summary and conclusions of the work.

%-----------------------------------------------------------------
\section{Methodology} \label{Sec:Simulation&Sample}
%-----------------------------------------------------------------

\subsection{Overview of the Auriga cosmological simulations} \label{Sec:overview_simu}

We make use of the cosmological gravo-magnetohydrodynamic `zoom-in' Auriga simulations \citep{Grand_2017,Grand_2024}. These simulations were designed to understand the formation and evolution of Milky Way-type spiral galaxies. We use $30$ haloes with halo masses at $z=0$ of $M_{\rm 200}$ = $1-2 \times 10^{12}\rm{M}_{\odot}$, which were first presented in \cite{Grand_2017}. These are selected from a parent dark matter-only counterpart to the Eagle simulation, with a comoving box size of $100 \, \rm{cMpc}$. The $30$ haloes were selected randomly from a sample of $174$ candidates that meet the conditions of (a) being in a mass range of $1 < M_{200}/10^{12} M_{\odot} < 2$ and (b) being relatively isolated at $z=0$. The isolation criteria considers haloes that have a mass at least $3$ per cent greater than that of the main halo and are located at a distance of at least $9$ times their $R_{\rm vir}$ from the main halo. The simulations were performed using the {\sc arepo} moving-mesh code \citep{Springel_2010, Pakmor_2015}, following the coupled evolution of the dark matter and gas components, and implement a galaxy formation model described in \cite{Grand_2017}. The cosmological parameters that the simulations adopt are $\Omega_{\mathrm m} = 0.307$, $\Omega_{\mathrm b} = 0.048$, $\Omega_{\Lambda} = 0.693$ and a Hubble constant of $H_0 = 100 \, h \, \rm{km} \, \rm{s}^{-1} \, \rm{Mpc}^{-1}$, where $h = 0.6777$, taken from \cite{Planck2013_2014}.

In this work we use simulations with a baryonic mass resolution of $5 \times 10^{4} \, \rm{M}_{\odot}$ and a dark matter mass resolution of $4 \times 10^{5} \, \rm{M}_{\odot}$. The softening length of the star particles is approximately $375 \, \rm{pc}$ and for the gas cell is scaled by the mean radius of the cell.
 
%-----------------------------------------------------------------
\subsection{Identifying the barred galaxies} \label{Sec:galaxies_properties}

\begin{table} 
    \caption{Properties at $z=0$ of the galaxies in the sample: galaxy simulation name, stellar and gas mass inside $10 \, \rm kpc$ radius, bar strength, and bar radius calculated using the mode $A_2(r)$.}
        \label{tab:properties_z_0}    
        \centering
            \begin{tabular}{ccccc}
                \hline 
                Galaxy & $M_{*}$ & $M_{\rm{gas}}$ & $A_2^{\rm max}$ & $R_{\rm bar}$ \\
                 - & $10^{10} M_{\odot}$ & $10^{10} M_{\odot}$ & - & kpc \\
                \hline 
                Au09 & 5.23 & 0.43 & 0.48 & 3.5 \\ 
                Au10 & 5.57 & 1.01 & 0.52 & 3.7 \\ 
                Au11 & 4.33 & 0.78 & 0.46 & 7.7 \\ 
                Au13 & 5.42 & 0.80 & 0.40 & 4.5 \\ 
                Au17 & 6.72 & 0.77 & 0.47 & 4.7 \\ 
                Au18 & 6.48 & 0.26 & 0.36 & 4.5 \\ 
                Au22 & 5.67 & 0.51 & 0.39 & 3.7 \\ 
                Au23 & 6.67 & 0.47 & 0.38 & 4.9 \\ 
                Au26 & 9.77 & 0.82 & 0.42 & 3.7 \\ 
                \hline
            \end{tabular}
\end{table} 

We are interested in conducting an evolutionary study of the various properties that characterise the b/p structures in the galaxies of the Auriga cosmological simulations. For this reason, it is crucial to detect when these structures first emerge and to trace their evolution throughout the simulation timeline. In order to identify the b/p's we first need to know which galaxies host a bar (see also \citealt{Fragkoudi_2024} for the barred galaxy fraction in the extended Auriga suite of simulations, a different sample from the one used in this work). To do that, we perform a Fourier decomposition of the face-on ($xy$ plane) density distribution of the stellar disc of the $30$ simulated galaxies and calculate the modes \citep{Athanassoula_2013},
\begin{equation}
    \begin{aligned}
        &a_{m}(R)=\sum_{i} m_{i} \cos \left(m \theta_{i}\right), m \geq 0, \\
        &b_{m}(R)=\sum_{i} m_{i} \sin \left(m \theta_{i}\right), m>0,
    \end{aligned}
\end{equation} 
where the sum is over all the disc particles, defined as the ones with $|z|<5\, \rm kpc$\footnote{We note that the cut employed for the $z$-coordinate here is different from the one used in \cite{Fragkoudi_2024}, who used a cut of $|z|<0.8 \, \rm kcp$. The larger cut employed here will lead to lower peaks in the $A_{2}$ profile, which leads to slightly younger bar ages in this study, as compared to \cite{Fragkoudi_2024}, i.e. by about $\sim 11\%$. However, we note that this choice does not qualitatively affect the results of our study.} and projected radius inside $10\, \rm kpc$, $m_i$ is the mass of particle $i$ and $\theta_i$ is the azimuthal angle. We perform the calculation by taking radial bins of $\Delta r = 0.2\, \rm kpc$ width from the centre of each galaxy. The strength of the bar is defined from the $m=2$ mode of the Fourier decomposition 
\begin{equation}
    A_2 = \frac{\sqrt{a_2^{2}+b_2^{2}}}{a_0}.
\end{equation}
From this calculation we can derive the strength of the bar for each snapshot of the simulation, taking the maximum of the $m=2$ mode, defined as $A_{2,\rm max}$. We consider that a bar is formed when the maximum strength reaches a threshold of $0.25$ and continues over that value thereafter; as a result of this analysis we find $18$ galaxies having a bar at $z=0$. 

To calculate the sizes of the bars, $R_{\rm bar}$, we take the radius where the radial profile of $A_2$ has a drop of $0.35 \, A_{2,\rm max}$ (i.e. $A_2$ reaches a value of $0.65 \, A_{2,\rm max}$ after its maximum), after reaching its maximum. 

%-----------------------------------------------------------------
\subsection{Method to detect b/p bulges} \label{Sec:Method}

\begin{figure}
    \centering
    \includegraphics[width=1\columnwidth]{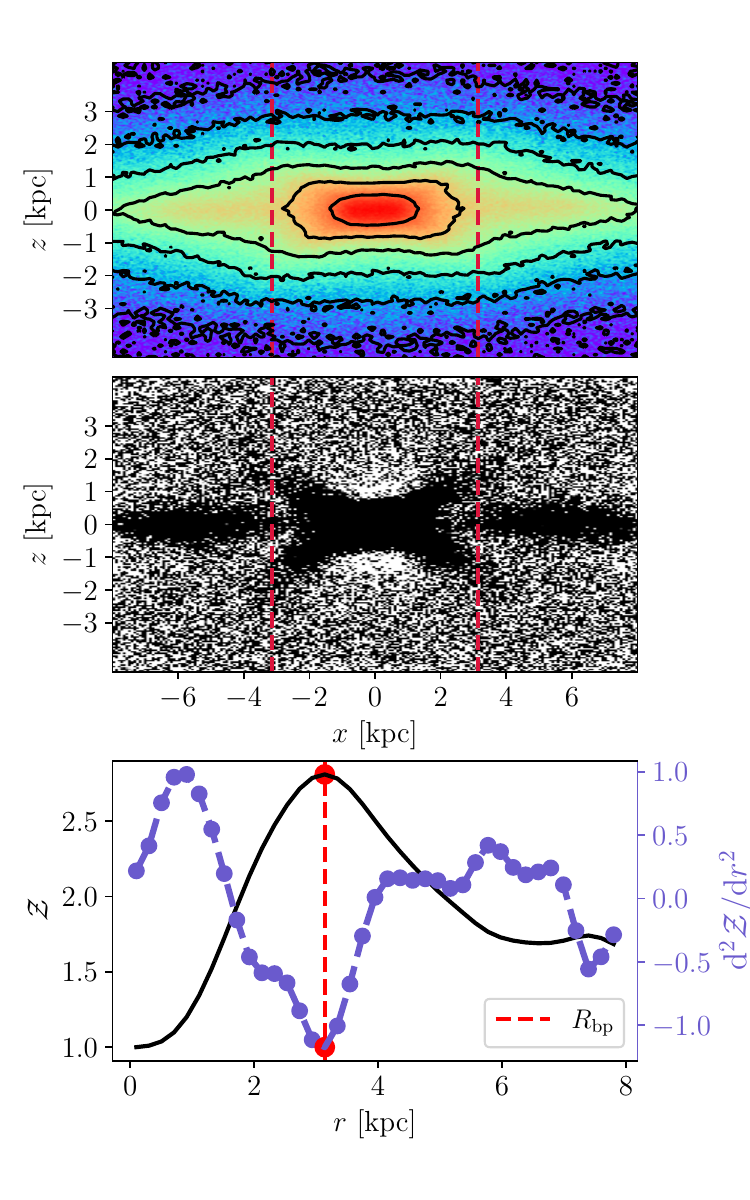}

    \caption{\textit{Upper panel}: Stellar surface density in the edge-on projection ($xz$ plane) for the simulated galaxy Au23 at $z=0$. \textit{Middle panel}: Unsharp mask of the surface density in the edge-on projection. \textit{Lower panel}: Radial profile of the median absolute value of the stellar heights, normalised by the central value evaluated at $z=0$, $\varcal{Z}$ (solid black line), and the second derivative of $\varcal{Z}$ (dashed violet curve with dots). The vertical dashed lines in all panels indicate the size of the b/p structure, $R_{\rm bp}$.}
    \label{fig:Au23_gradient}
\end{figure}

Once we have the sample of barred galaxies at $z=0$, we conduct a visual analysis of the stellar density distributions and unsharp masks\footnote{Unsharp masking refers to a technical process applied to an image where blurred components are removed, therefore enhancing the features present in the original version of the image. For this, we use the unsharp masking filter from the {\sc scikit-image python} library. } of the edge-on projection; we show the simulated galaxy Au23 as an example in the upper and middle panels of Fig. \ref{fig:Au23_gradient}. We identify $9$ galaxies with a b/p structure: $6$ of them were analysed in a previous work \citep{Blazquez-Calero_2020}, but we also find $3$ more galaxies hosting a b/p bulge. These $9$ galaxies constitute our b/p bulge sample, and their properties are listed in Table \ref{tab:properties_z_0}.

Various methods have been employed previously in the literature for deriving the presence and strength of b/p's; for example, some studies have used the kurtosis of the vertical velocity or the absolute value of the z-component of stars \citep{Debattista_2005, Martinez-ValpuestaAthanassoula_2008,Fragkoudi_2017,Anderson_2023, Ghosh_2024}.
\begin{figure}
    \centering
    \includegraphics[width=1\columnwidth]{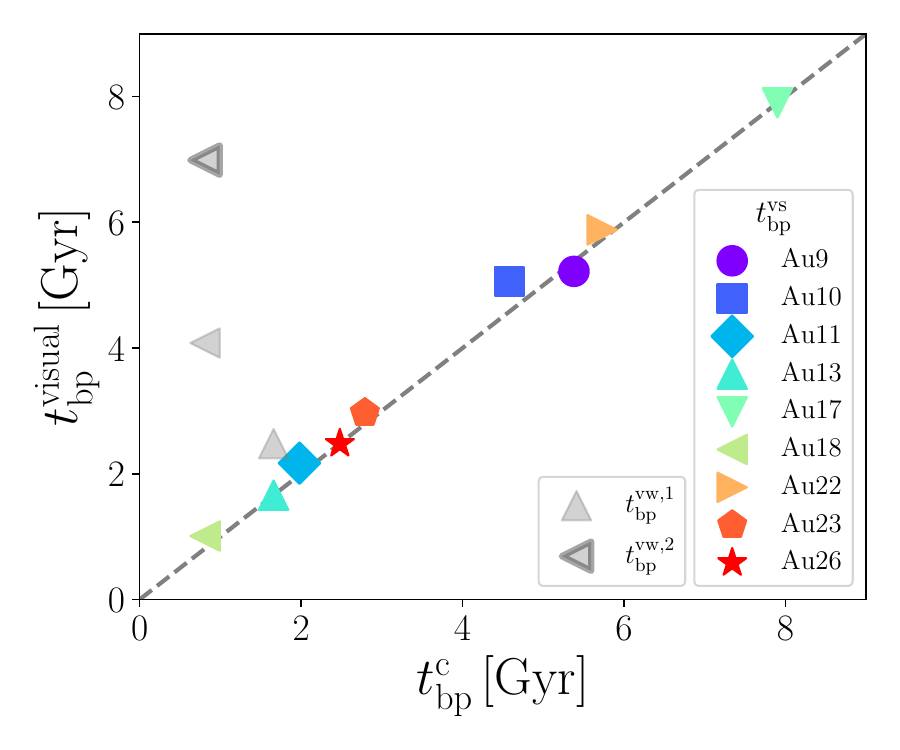}
    \caption{Correlation between two of the definitions for the formation time of the b/p bulge: the visual ones (visual strong in colours and the visual weak in grey)} versus the one obtained from the peak-detection code. The diagonal dashed line corresponds to the identity line. All the times are lookback times in Gyr.
    \label{fig:tvs_tcode}
\end{figure}
\begin{figure*}
    \centering
    \includegraphics[width=\textwidth]{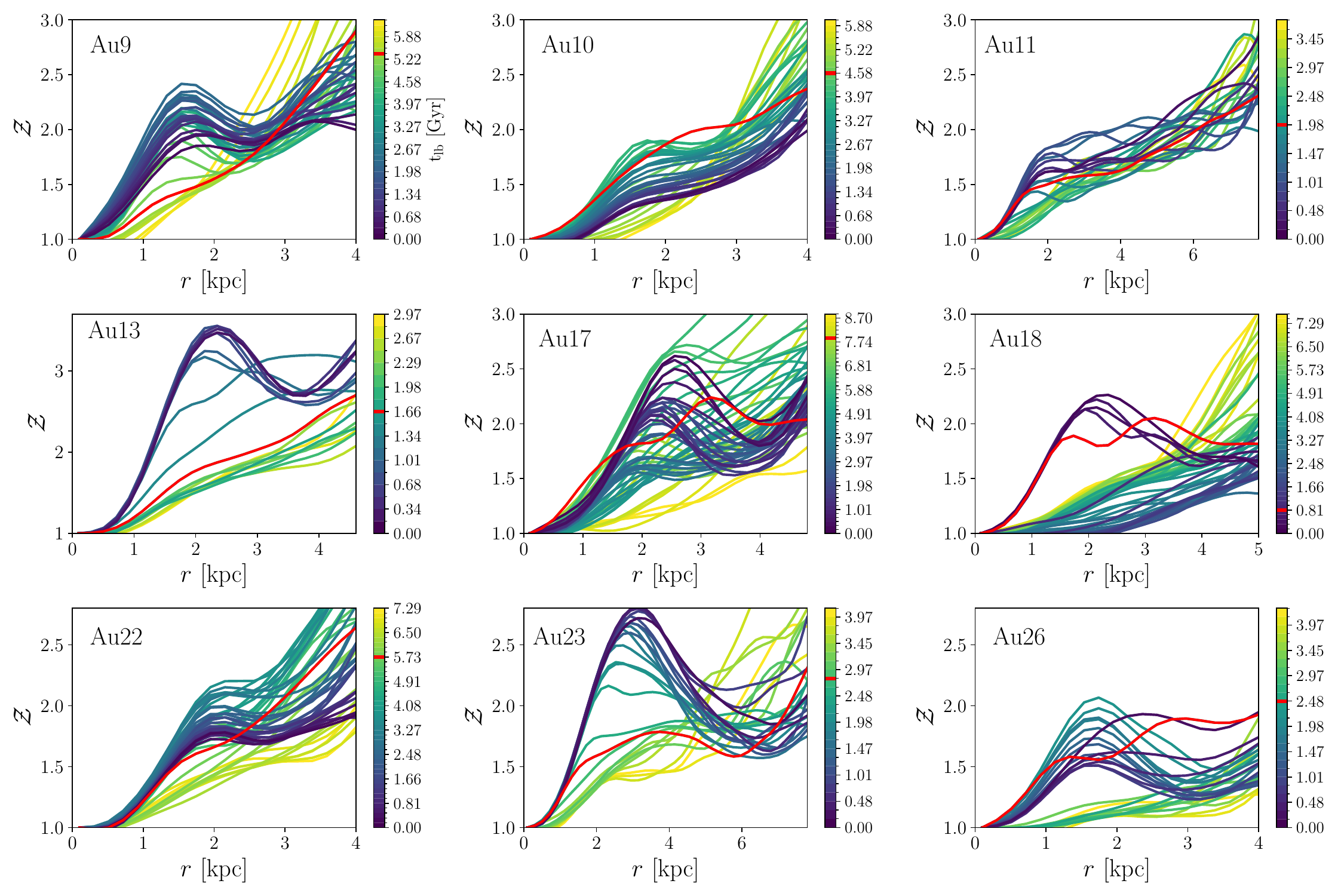}
    \caption{Radial (cylindrical) profiles of the median of the absolute value of the $z$-positions of stars normalised by the central value.
    The different colours are the lookback time measured since bar formation ($t_{\rm bar}$) until $t_{\rm lb} = 0 \, \rm Gyr$. The code peak-detected time formation of the b/p bulge ($t_{\rm bp}^{\rm c}$) is indicated with a red line.}
    \label{fig:bp_strength_radius}
\end{figure*}
For this work, we tested various methods and developed an automatised algorithm to detect the presence of a b/p and its strength which gives us a good match to the visual classification of b/p bulges (based on unsharp masking). We describe the steps of the procedure as follows:
\begin{enumerate}
    \item In the rotating reference system with the bar, we isolate the bar structure of each galaxy and follow its formation with time. For this purpose, we select the stars inside the bar and close to the plane, i.e. with the absolute value of the \textit{y} and \textit{z}-coordinates satisfying the conditions $|y| < 0.8 \, \rm kpc$, $|z| < 2 \, \rm kpc$, and $R < R_{\rm bar}^{z=0}$, where $R$ is the cylindrical radius and $R_{\rm bar}^{z=0}$ refers to the bar length at $z=0$. 
    
    \item We calculate the median of the absolute value of the height of the star particles above and below the disc, $|z|_{\rm median}$, by using radial bins of $0.2 \, \rm kpc$ and then normalize the profile to the value of the median of $|z|$ in the centre of the galaxy ($|z|_{\rm median}(R=0)$). From now on, we define
    \begin{equation} 
        \varcal{Z} = \frac{|z|_{\rm median}}{|z|_{\rm median}(R=0)}.
    \end{equation}
    We calculate $\varcal{Z}(r)$ for the $9$ galaxies at every snapshot. 
    A peak in the profiles is associated with the signal of the b/p bulge; an example of this is shown for Au23 in the lower panel of Fig. \ref{fig:Au23_gradient}, where we show $\varcal{Z}(r)$ on the left $y$-axis (solid black line).
       
    \item We implement a method that involves detecting this peak, i.e. a local maximum in $\varcal{Z}$, by requiring to detect a minimum negative value in the second derivative of $\varcal{Z}$ ($\mathrm{d} ^2 \varcal{Z}(r)/\mathrm{d}r^2$). To do this, we exclude the inner and outer radial region taking $0.15 \, R_{\rm bar}^{z=0} < r < 0.75 \, R_{\rm bar}^{z=0}$. In order to facilitate peak detection we have smoothed the profiles with a Butterworth filter \citep{Butterworth1930}, using the {\sc signal} module from {\sc scipy python} library. In the lower panel of Fig. \ref{fig:Au23_gradient}, we illustrate $\mathrm{d} ^2 \varcal{Z}(r)/\mathrm{d}r^2$ at $z=0$ for the galaxy Au23, where the peak is highlighted by a red dot. We use the radial position of the peak as a definition for the size of the b/p bulge, $R_{\rm bp}$, indicated by the vertical red dashed line (see in appendix \ref{App:bp_z0} Fig. \ref{fig:density_unsharp_plots} for the rest of the galaxies).

    \item We define the strength of the b/p bulge, $BP_{\mathrm{\mathrm{strength}}}$, as the value of $\varcal{Z}$ at $R_{\rm bp}$, and trace this over time. We set the b/p' strength to zero if a peak is not detected in the profile of $\varcal{Z}$.

    \item To determine the formation lookback time of the b/p bulges, we refer to Fig. \ref{fig:nabla_2_min} in Appendix \ref{App:bp_detection}, which shows the evolution of the minimum value of the second derivative of $\varcal{Z}$, $\mathrm{d} ^2 \varcal{Z}(r)/\mathrm{d}r^2$, from the formation lookback time of the bar, $t_{\rm bar}$, to the present ($t_{\rm lb} = 0 \, \rm{Gyr}$) for each galaxy. When this parameter drops below a threshold\footnote{The threshold was determined by visually inspecting when a b/p appears in edge-on images of the galaxy.} of $\mathrm{d} ^2 \varcal{Z}(r)/\mathrm{d}r^2 < -0.51$, and remains below it consistently (for at least 5 snapshots), we define this instant as marking the formation time of the b/p bulge, denoted as $t_{\rm bp}^{\rm c}$\footnote{Throughout the paper, when we talk about formation times we are referring to lookback times.}. 

\end{enumerate}
\begin{table}
\caption{Lookback times in Gyr for the formation of the stellar bar, $t_{\rm{lb,bar}}$, and the b/p bulge for galaxies in our sample. 
Four formation times are defined for the b/p bulge: 
the times identified by visual inspection when it can be seen as a weak one ($t_{\rm{bp}}^{\rm vw,1}$), 
a very weak one ($t_{\rm{bp}}^{\rm vw,2}$) (only defined for Au18), 
a strong b/p ($t_{\rm{bp}}^{\rm vs}$), 
and the time determined by the peak-detection code ($t_{\rm{bp}}^{\rm c}$).}
    \label{tab:times_bp}
    \centering
    \begin{tabular}{ccccccc}
        \hline 
        Galaxy & 
        $\begin{array}{c} t_{\rm{bar}}         \end{array}$ &
        $\begin{array}{c} t_{\rm{bp}}^{\rm vw,1} \end{array}$ &
        $\begin{array}{c} t_{\rm{bp}}^{\rm vw,2} \end{array}$ &
        $\begin{array}{c} t_{\rm{bp}}^{\rm vs} \end{array}$ & 
        $\begin{array}{c} t_{\rm{bp}}^{\rm c} \end{array}$ \\
        \hline 
        Au09 & $6.50$ & -      & -      & $5.22$ & $5.38$ \\
        Au10 & $6.17$ & -      & -      & $5.06$ & $4.58$ \\
        Au11 & $3.96$ & -      & -      & $2.17$ & $1.98$ \\
        Au13 & $3.27$ & $2.48$ & -      & $1.66$ & $1.66$ \\
        Au17 & $9.16$ & -      & -      & $7.90$ & $7.90$ \\
        Au18 & $7.90$ & $4.08$ & $6.99$ & $1.01$ & $0.81$ \\
        Au22 & $7.62$ & -      & -      & $5.88$ & $5.73$ \\
        Au23 & $4.42$ & -      & -      & $2.97$ & $2.79$ \\
        Au26 & $3.62$ & -      & -      & $2.48$ & $2.48$ \\
        \hline
    \end{tabular}
\end{table} 
To evaluate the accuracy of the detected formation times using our method, we carry out a visual inspection for all galaxies across all times after bar formation ($0 \, \rm Gyr < t_{\rm lb} < t_{\rm bar}$), examining both the stellar surface density and the unsharp mask in the edge-on projection. As a result, we visually identify the moment when the b/p bulge clearly emerges (appearing as a \textit{strong} feature in the edge-on projections), calling it: $t_{\rm{bp}}^{\rm vs}$. Fig. \ref{fig:tvs_tcode} shows a comparison between the time detected by the code (peak-detection method, $t_{\rm{bp}}^{\rm c}$) and the time at which strong b/p's are identified visually ($t_{\rm{bp}}^{\rm vs}$ in colours). Most galaxies fall close to the 1-to-1 line, suggesting a good agreement. During the visual analysis, we find that for two of the galaxies explored, they show signs of \textit{weak} b/p structures, which are not detected by our algorithm. These appear for a short period of time and subsequently are no longer detectable. We refer to these earlier detections of a weak b/p signal as $t_{\rm{bp}}^{\rm vw,1}$.  A noteworthy case is that of Au18, where we find two weak signals of b/p's at two distinct times: the first at $t_{\rm lb} \sim 4 \, \rm Gyr$ and the second at $t_{\rm lb} \sim 7 \, \rm Gyr$, meaning $3 \, \rm Gyr$ and $6 \, \rm Gyr$ before the strong b/p bulge is picked up by our algorithm (see Appendix \ref{App:bp_weak}). We define this earlier time, associated with the first appearance of a weak b/p structure, as $t_{\rm{bp}}^{\rm vw,2}$. However, we find this structure to be short-lived, appearing for a period of time of roughly $\sim 0.5 \, \rm Gyr$ before dissolving. Table \ref{tab:times_bp} lists the values for the different definitions of the b/p formation time for each galaxy.

For the galaxies in which we do not list a $t_{\rm{bp}}^{\rm vw,1}$ value, we find that there can be a weak b/p signal approximately $1-2$ snapshots before the strong b/p forms. This gives us the approximate time it takes for the galaxy to go from having no detectable b/p to a strong b/p, which is approximately $\sim 160 \, \rm Myr$.

%-----------------------------------------------------------------
\section{B/P formation} \label{Sec:bp_formation}

\begin{figure*}
    \centering
    \includegraphics[width=2\columnwidth]{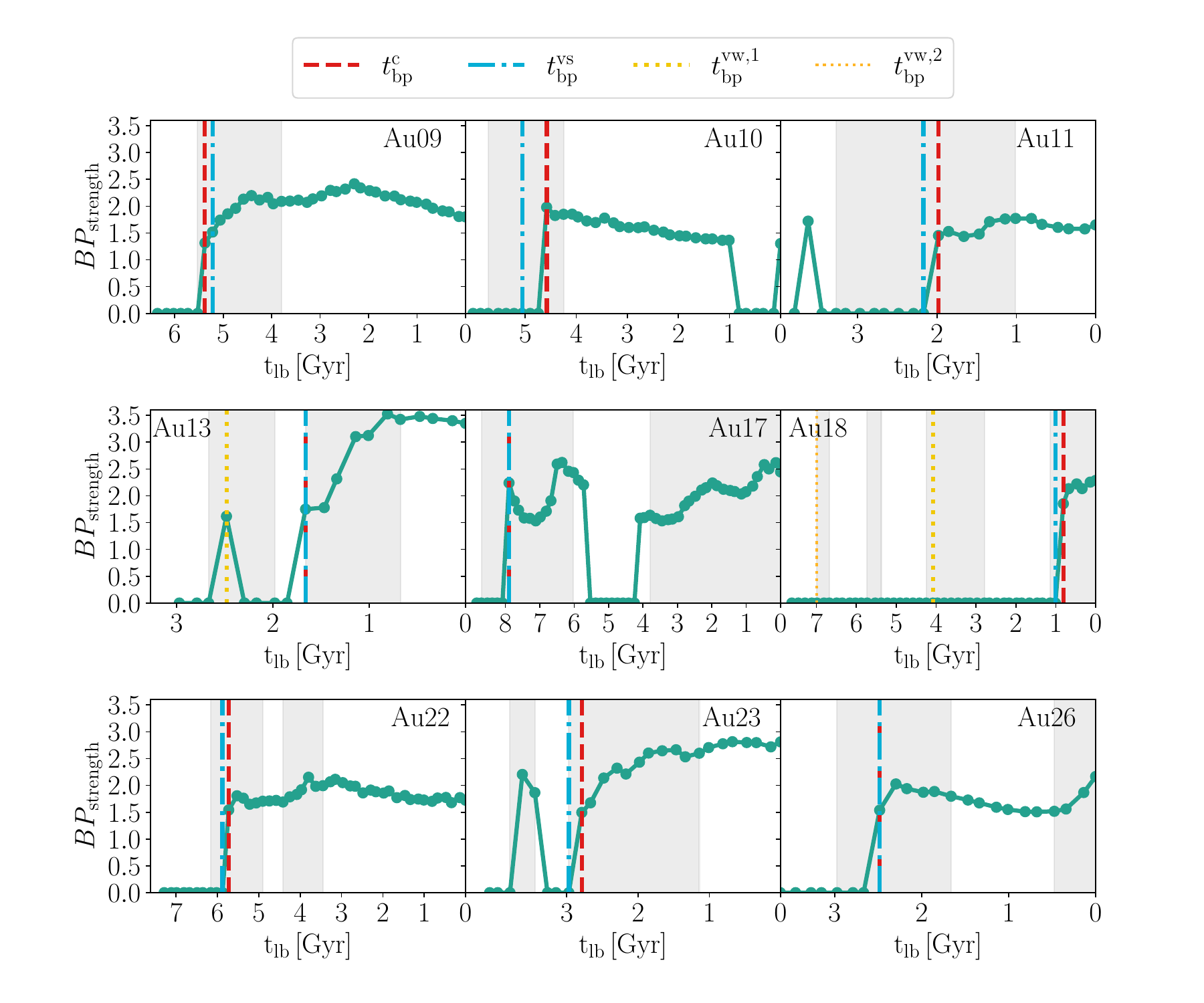}
    \caption{Evolution of the strength of the b/p bulge, obtained from the peak of the distribution of $\varcal{Z}$ as a function of the cylindrical radius $r$. The profiles start at the lookback time $t_{\rm{bar}}$ of each galaxy. The dashed line is the lookback time formation of the b/p bulge according to the peak-detection code ($t_{\rm bp}^{\rm c}$), the dashed-dotted line is the lookback time formation of the visually strong b/p bulge ($t_{\rm bp}^{\rm vs}$), the thick dotted line is the lookback time formation of the visually weak b/p bulge ($t_{\rm bp}^{\rm vw,1}$), and the thin dotted line is the lookback time formation of the visually very weak b/p bulge ($t_{\rm bp}^{\rm vw,2}$), defined for Au18. The shaded regions indicate the lookback times at which the bar is visually detected to be buckling (see Sec. \ref{Sec:buckling_episodes}).}
    \label{fig:bp_strength_time}
\end{figure*} 
\begin{figure}
    \centering
    \includegraphics[width=1\columnwidth]{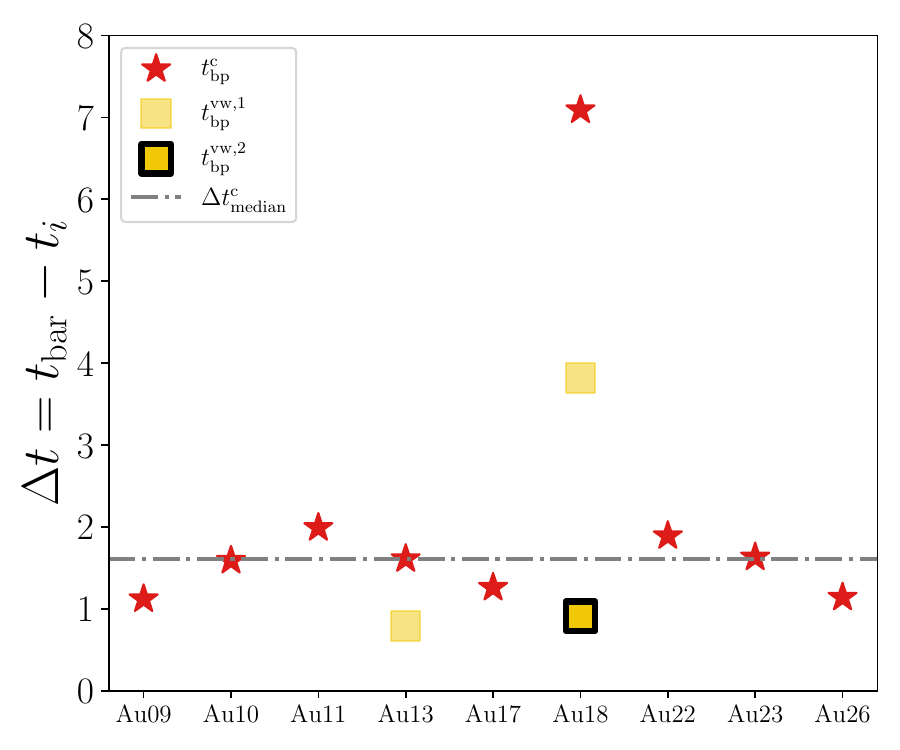}
        \caption{Time interval between the formation time of the b/p bulge with respect to the formation time of the formation time of the bar as determined by visual inspection of a weak structure (square), a very weak structure for Au18 (square with thick lines), and according to the constructed code (star). The horizontal dashed-dotted line correspond to the median based on the code detection method ($\Delta t^{\rm c}_{\rm median}$).}
    \label{fig:delta_t}
\end{figure}
%-----------------------------------------------------------------
In this Section, we explore the formation time of the b/p's and how common they are in these simulations.
As we mentioned before, to determine the formation time of the b/p structure, we computed the radial profile of $\varcal{Z}$ for all time steps after $t_{\rm bar}$, as shown in Fig. \ref{fig:bp_strength_radius}, with different colours representing times since bar formation in lookback time. The profiles differ across galaxies, some showing more pronounced peaks than others and, generally, all profiles show an increase towards larger radii, a trend not observed in isolated galaxy simulations (see e.g. \citealt{Ghosh_2024}). This flaring effect is a consequence of star formation occurring in a flaring gaseous disc \citep{Grand_2017} combined with interactions between the stellar disc and satellites, which heats up the outer disc \citep{Grand_2016}. Initially, right after bar formation, we can see that there is no visible peak in $\varcal{Z}$, indicating the absence of a b/p structure. After some time, we observe how the curves begin to display more defined peaks, indicating the emergence of the b/p bulge. The red curve in Fig.  \ref{fig:bp_strength_radius} represents the $\varcal{Z}$ profile at $t_{\rm bp}^{\rm c}$ for each galaxy (see Sec. \ref{Sec:Method}). Once the formation time of each b/p bulge is established we can track its evolution over time. To analyse this we can refer to Fig. \ref{fig:bp_strength_time} that shows the strength of the b/p bulge as a function of lookback time. Overlaid are the different definitions of the b/p formation times, and the shaded regions correspond to time intervals where buckling episodes were detected, which we will discuss further in Sec. \ref{Sec:buckling_episodes}.

Several interesting insights emerge from this analysis. In general, some galaxies show a clear increase in their b/p strength over time, as seen in Au13 and Au23. Notably, the code successfully detects the presence of a b/p when it is weakly observed in Au13. However, this is not always the case, as both Au11 and Au23 show a signal in their b/p profiles, but it is not associated with the actual presence of a b/p structure. Other galaxies exhibit an initial growth in strength during the first few Gyr of evolution, followed by a slight decrease, eventually reaching similar values at $z=0$  to those observed when the b/p first emerged. This behaviour is seen in the profiles of Au09, Au11, and Au22. Additionally, we observe some particular behaviours unique to each galaxy. In the case of Au10, a b/p structure formed approximately $5 \, \rm Gyr$ ago, with its strength gradually decreasing over time, disappearing almost completely in the last gigayear of evolution, only to be detected again at $z=0$. Fig. \ref{fig:Au10_desapears} in Appendix \ref{App:bp_fading} displays the stellar density distribution and unsharp mask of the edge-on projection of Au10 during the period when the b/p is no longer detected. There is agreement between the code, which does not detect the structure, and the visual inspection, where no clear b/p is observed, while at $z=0$ the b/p becomes distinguishable once again, leading to its detection by the code. This raises several questions: is the structure present but so weak that it cannot be detected, or does it truly vanish? If it remains but is too faint, should it still be considered a b/p, given that it would likely not be detected observationally? Moreover, why does the structure disappear in the first place? Among the other galaxies, Au17 presents another case of b/p absence at some point during its evolution. Observing its strength evolution, the galaxy experiences a significant increase in strength early on, followed by a decrease starting at around $t_{\rm lb} \sim 6 \, \rm Gyr$. The b/p then disappears, re-emerging $2 \, \rm Gyr$ later with much weaker strength. From that point, its strength increases again, though with some fluctuations, recovering significantly by $z=0$. Fig. \ref{fig:Au17_desapears} shows the stellar density distribution and the unsharp mask of the edge-on projection for Au17 during the period of time where the b/p bulge disappears. Additionally, Au26 shows a general decrease in strength throughout most of its evolution, although it seems to recover in the last $0.5 \, \rm Gyr$ lookback time. Finally, the b/p bulge in Au18 has had limited time to evolve, as it formed in the last gigayear of the simulation. However, this galaxy is notable for being the only one with two separate times, $t_{\rm{bp}}^{\rm vw,1}$ and $t_{\rm{bp}}^{\rm vw,2}$, where a structure resembling a b/p is observed in the edge-on projections. Both instances are weak and short-lived, visible only for a couple of snapshots.

This analysis shows that b/p bulges in galaxies follow different evolutionary patterns. Some galaxies experience growth in b/p strength and then a decrease, while others go through phases of weakening or even temporary disappearance, followed by a recovery, suggesting that the formation and evolution of b/p structures are not uniform. The peak-detection method generally agrees with visual observations, though there are cases where it identifies weak signals that do not clearly correspond to a visible b/p. This shows the complexity of detecting such structures within this type of simulation and the importance of combining both automated techniques and visual inspection to accurately track the b/p structures' formation and evolution.

In Fig. \ref{fig:delta_t} we investigate whether there is a characteristic timescale between the formation of a bar and the formation of a b/p structure in galaxies that exhibit this feature.
We show $\Delta t$ for each galaxy, which is defined as the time interval between the formation of the bar ($t_{\rm bar}$) and the formation of the b/p structure. The latter is measured through different methods: the peak-detection method ($t_{\rm bp}^{\rm c}$, marked with stars), weak visual detection ($t_{\rm bp}^{\rm vw,1}$, marked with squares) and, for Au18, the very weak visual detection ($t_{\rm bp}^{\rm vw,2}$, represented by a square with a thicker border). We find that b/p's in the Auriga galaxies tend to form shortly after bar formation. The median time that elapses between the formation of the bar and the formation of the b/p is
$\Delta t^{\rm c}_{\rm median} = 1.6 \, \rm Gyr$ (horizontal dashed-dotted line). It is interesting to note that in the case of Au18, where a strong b/p bulge forms very late compared to when the bar forms -- specifically $\sim 7 \, \rm Gyr$ after $t_{\rm bar}$ -- there are previous signatures of weak b/p structures, detected at $\sim 1$ and $3.8 \, \rm Gyr$ after the formation of the bar. Our findings possibly hint at a characteristic timescale for b/p formation for galaxies in a cosmological setting. However,  we emphasize our small sample size, and the fact that this timescale will likely depend on details of the galaxy formation model.

%-----------------------------------------------------------------
\subsection{b/p fraction} \label{Sec:bp_fraction}
%-----------------------------------------------------------------
\begin{figure*}
    \centering
    \includegraphics[width=2\columnwidth]{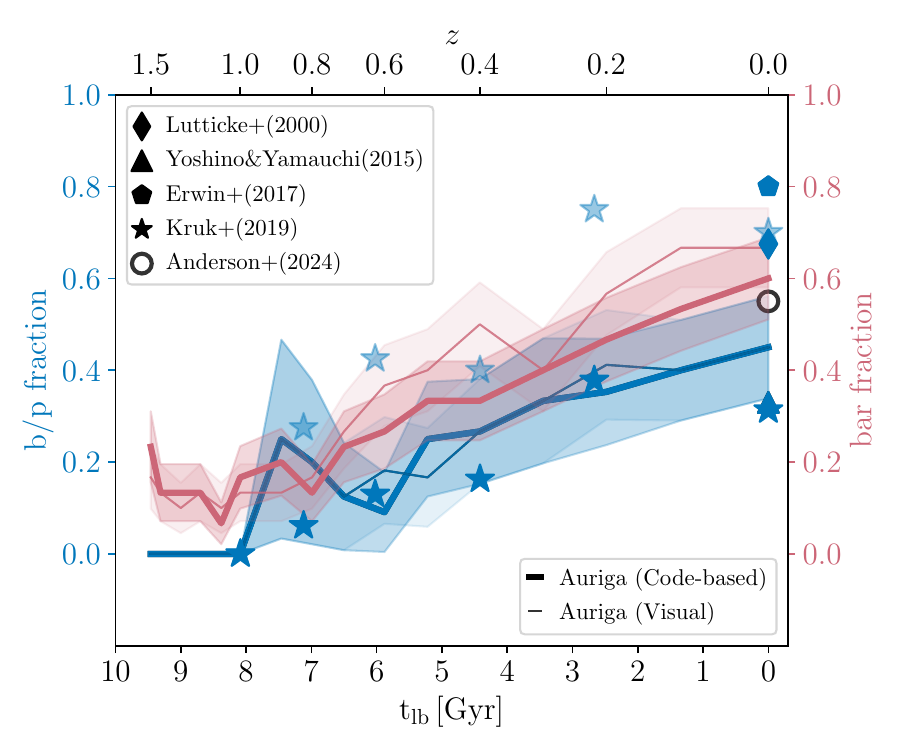}
    \caption{B/p fraction (left y-axis in blue) and bar fraction (right y-axis in pink) as function of redshift, considering the $30$ galaxies in the simulation. Two methods were used to obtain the fractions in each case: a code-based count method (thick lines) based on the strength of each component and a visual count method (thin lines). The shaded regions represent the uncertainties in the fractions, calculated assuming binomial statistics. The symbols represent the results from observations: \citet{Lutticke_2000} (diamond), \citet{Yoshino_Yamauchi_2015} (triangle), \citet{Erwin_2017} (pentagon), and the observed (dark blue stars) and the corrected (light blue stars) fractions by \citet{Kruk_2019}; and from a cosmological simulation: \citet{Anderson_2023} (empty circle).}
    \label{fig:bp_fraction}
\end{figure*}
Although b/p bulges are the vertically extended part of evolved bars, not all barred galaxies contain a b/p structure. This raises some interesting questions: What is the fraction of barred galaxies that contain a b/p bulge? Why do some barred galaxies not form a b/p bulge? We aim to address these questions by analysing the barred galaxies within the Auriga simulations\footnote{Referring to the $30$ original Auriga galaxies presented in \cite{Grand_2017}.} over time and discuss some observational findings.

To explore the fraction of b/p's we first need to find the galaxies that have bars. We therefore first calculate the bar fraction by applying two identification methods, one based on visual classification (visual method) and one using the $A_{\rm max,2}$ parameter (code-based method; see Sec. \ref{Sec:galaxies_properties} for more details).
These fractions are shown in Fig. \ref{fig:bp_fraction} in pink (see the right y-axis), where the thick line corresponds to the code-based method and the thin line to the visual method. The two methods are in good agreement and show a decreasing trend with redshift. For most of the evolution, the visual method gives a slightly higher fraction than the code-based method; this is because the code-based method is sensitive to both the threshold used to determine the presence of the bar ($A_{2,\rm max} > 0.25$) and to the cut used to select the stars in the disc (i.e. $|z| < 5 \, \rm{kpc}$). On the other hand, at higher redshifts the galaxies might have large $A_{\rm max,2}$ values because of mergers and interactions, and not due to the presence of a bar. After obtaining the barred galaxy fractions for the different redshifts, we proceeded to calculate the b/p fraction for the same redshifts. We again used two methods to count the galaxies with a b/p bulge: the peak-detection method (code-based) approach relied on the b/p strength (see Fig. \ref{fig:bp_strength_time}) and the visual method (see Sec. \ref{Sec:Method}). The b/p fraction is obtained by dividing that number by the number of barred galaxies (as identified by the visual method). 

We show these results in Fig. \ref{fig:bp_fraction} in blue (see the left y-axis). We find that at $z=0$, $66$ per cent of the galaxies host a bar ($60$ per cent if we use the code-based method), and almost half of these barred galaxies, $45$ per cent, host a b/p structure. 
Both the bar and the b/p fraction decrease with redshift, and by $z \sim 1$ none of the galaxies exhibit a b/p bulge.
The shaded regions represent the uncertainties in the fractions, calculated assuming binomial statistics. These uncertainties are given by $\sigma^2 = f_{\rm bar}(1-f_{\rm bar})/N_{\rm all}$ for the bar fraction and $\sigma^2 = f_{\rm bp}(1-f_{\rm bp})/N_{\rm bars}$ for the b/p fraction, where $N_{\rm all}$ is the total number of galaxies and $N_{\rm bars}$ is the number of barred galaxies.  

In Fig. \ref{fig:bp_fraction} we also show the b/p fraction at $z=0$ from the IllsutrisTNG50 simulation from the work of \cite{Anderson_2023} and the observed b/p fraction from the works of \cite{Erwin_2017}, \cite{Lutticke_2000} and \cite{Yoshino_Yamauchi_2015}. In the first study, the authors obtain the b/p fraction from the visual inspection of the edge-on projection of barred galaxies at $z=0$. In the latter two works, the authors determine the fraction of b/p structures within a sample of edge-on galaxies (i.e. they obtain the fraction of b/p bulges over \emph{all} disc galaxies, rather than over \emph{barred} galaxies). To be able to compare their results to our b/p fraction (i.e. $N_{\rm b/p}/N_{\rm bar}$) we scale their fraction by a factor of $3/2$, to account for the fact that approximately two-thirds of disc galaxies in the local Universe host bars. 
Additionally, in Fig. \ref{fig:bp_fraction} we include the results of \cite{Kruk_2019}, who derive the temporal evolution of the b/p fraction up to $z \sim 1$. The authors give two sets of results: the observed b/p fraction (dark blue stars) and a corrected b/p fraction (light blue stars), in which they seek to account for observational biases (see Section \ref{Sec:Discussion} for more details). We note that, while in our study we are only exploring a subset of the galaxy population, specifically galaxies which end up within the stellar mass range, within $0.1 R_{\rm vir}$, of $10^{10.4} - 10^{11.1} \, M_{\odot}$ at $z=0$, our results are however broadly comparable with those from \cite{Kruk_2019}. In their work, the authors select galaxies with stellar masses $M_{\star} > 10^{10}\, M_{\odot}$ in the redshift range $0 \leq z \leq 1$; the Auriga galaxies we explore also have similar masses, i.e. $M_{\star} > 10^{10.4} \, M_{\odot}$, within the same redshift range. We do highlight however that the methods used for detecting bars and b/p bulges in our work might differ from those used in observational studies, while observational studies themselves also differ between them in the methods they employ (see Sec. \ref{Sec:Discussion} for more details). 

\begin{figure}
    \centering
    \includegraphics[width=1\columnwidth]{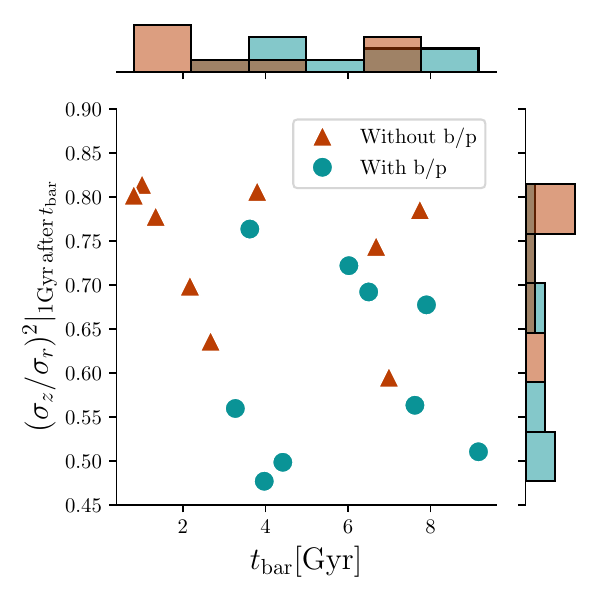}
    \caption{Ratio $(\sigma_{\rm z}/\sigma_{\rm r})^2$ for the barred galaxies evaluated $\sim 1 \, \rm{Gyr}$ after the formation of the bar, as a function of the bar age ($t_{\rm bar}$). The dots correspond to the barred galaxies hosting a b/p bulge and the triangles are the ones without it.}
    \label{fig:sigmazr}
\end{figure}
This leads us to other interesting question: why don't all barred galaxies form a b/p bulge? Theoretical studies and those using isolated N-body simulations have shown that b/p bulges are more likely to form in galaxies with kinematically cold discs \citep{Merritt_1994,Martinez-Valpuesta_2006}; this implies having a low ratio of the vertical to radial velocity dispersion of stars ($\sigma_z / \sigma_R$) (see Section \ref{Sec:Discussion} for more details). We calculate this ratio, as shown in Fig. \ref{fig:sigmazr}, for barred galaxies with (green dots) and without (red triangles) b/p bulges. This is calculated at $1 \, \mathrm{Gyr}$ after the formation of the bar. Considering that the b/p bulges form $\sim 1.6 \, \rm{Gyr}$ after the formation of the bar, we chose to evaluate the ratio $1 \, \rm{Gyr}$ after $t_{\mathrm{bar}}$ to analyse the values of the ratio before the formation of the b/p bulge but after the formation of the bar. We can see a general trend of galaxies without a b/p structure having younger bars and higher values of $(\sigma_z/\sigma_r)^2$, meaning that those galaxies possessing vertically hotter discs are prevented, or delayed, in forming a b/p bulge. There is one exception of a galaxy with an old bar and a low $\sigma_z/\sigma_r$ ratio, which does not host a b/p bulge, which indicates that there might be other factors that are important in b/p formation. On the other hand, the galaxies hosting a b/p bulge have bars with intermediate and old ages and with lower values of $(\sigma_z/\sigma_r)^2$, i.e. a colder disc. This highlights the importance of the ratio of vertical to radial stellar velocity dispersion in the formation of b/p's, while also suggesting that young bars are less likely to host a b/p (as is to be expected, since b/p's in general take some time to form after bar formation).

\section{Bar and b/p bulge evolution} \label{Sec:bar_bp_evolution}

\begin{figure} 
    \centering
    \includegraphics[width=1\columnwidth]{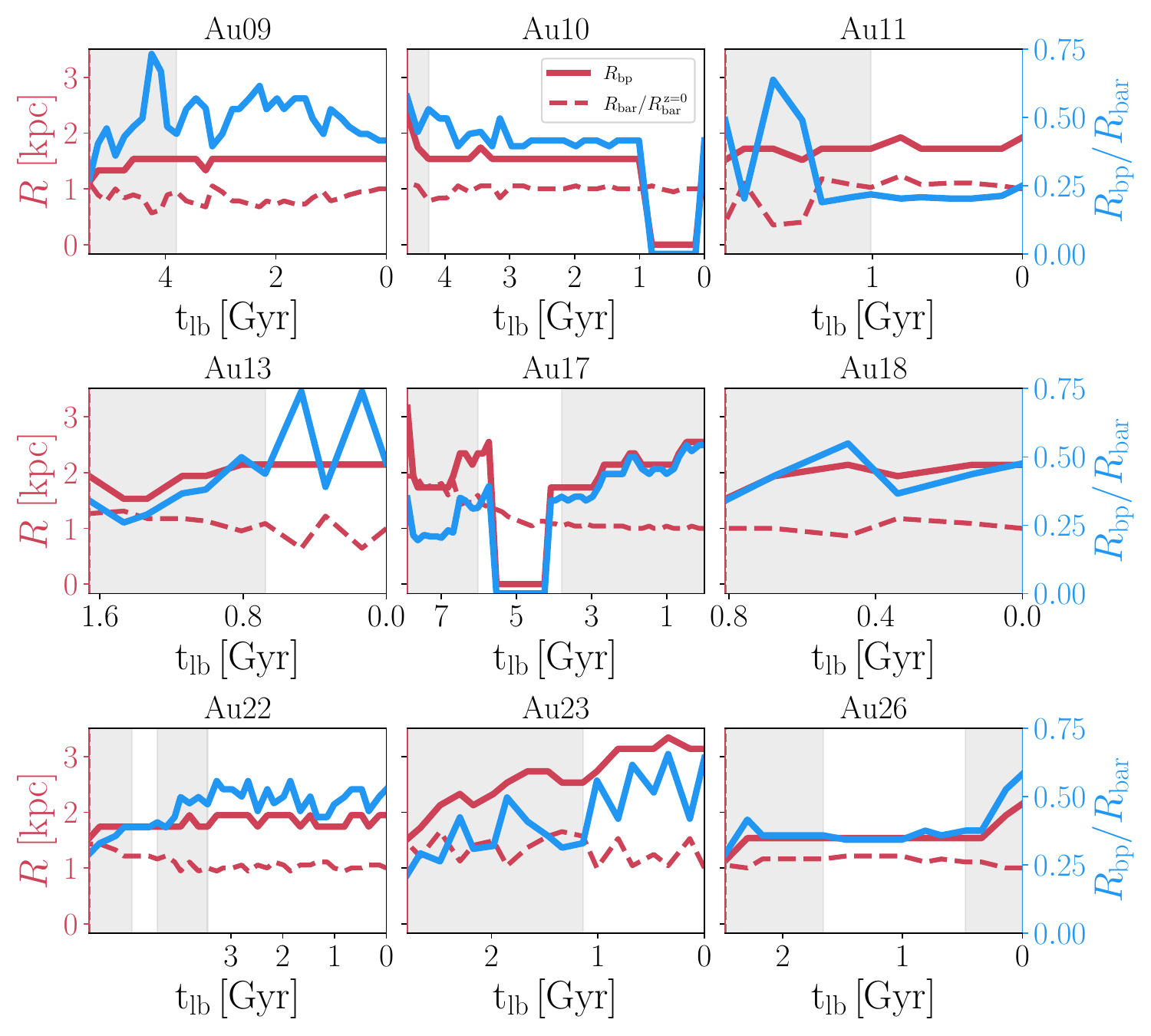}
    \caption{Left y-axis: b/p size as a function of lookback time (solid line) and bar size normalized by its size at $z=0$ (dashed line). Right y-axis: evolution in lookback time of the ratio between the b/p size and the bar size. The shaded regions indicate the lookback times at which the bar is visually detected to be buckling (see Sec. \ref{Sec:buckling_episodes}).}
    \label{fig:bp_size}
\end{figure}

In this section we explore the evolution of bar and b/p properties, and their interplay.
Fig. \ref{fig:bp_size} illustrates the b/p size as a function of lookback time for each galaxy, $R_{\rm bp}$ (see Sec. \ref{Sec:Method} for the definition), indicated on the left vertical axis with a red thick curve. We show the evolution of the bar size with a red dashed curve, normalized by the value at $z=0$, $R_{\rm bar}/R_{\rm bar}^{\, z=0}$. The evolution of these properties starts in the formation lookback time of the b/p, derived from the peak-detection method ($t_{\rm bp}^{\rm c}$). To understand how the evolution of the b/p bulge correlates with the overall evolution of the bar, the right vertical axis shows the ratio $R_{\rm bp}/R_{\rm bar}$, which represents the b/p size relative to the bar size. As mentioned in other figures, the shaded regions correspond to periods where buckling episodes occur, which will be explored further in Sec. \ref{Sec:buckling_episodes}. Regarding $R_{\rm bp}$, we observe different behaviours across galaxies. In general, the size tends to either increase or remain constant over time, as seen in Au09, Au11, Au18, Au22, and Au26. Au10 also shows stable evolution until the structure vanishes, leaving no associated size, a situation that also occurs with Au17. Galaxies such as Au13, Au17 (after reappearance), and Au23 display significant growth in b/p size over time. In terms of the link between $R_{\rm bp}$ and the $R_{\rm bar}$ with time evolution, we find a median value of $R_{\rm bp}/R_{\rm bar} = 0.43 \pm 0.06$.

\begin{figure} 
    \centering
    \includegraphics[width=1\columnwidth]{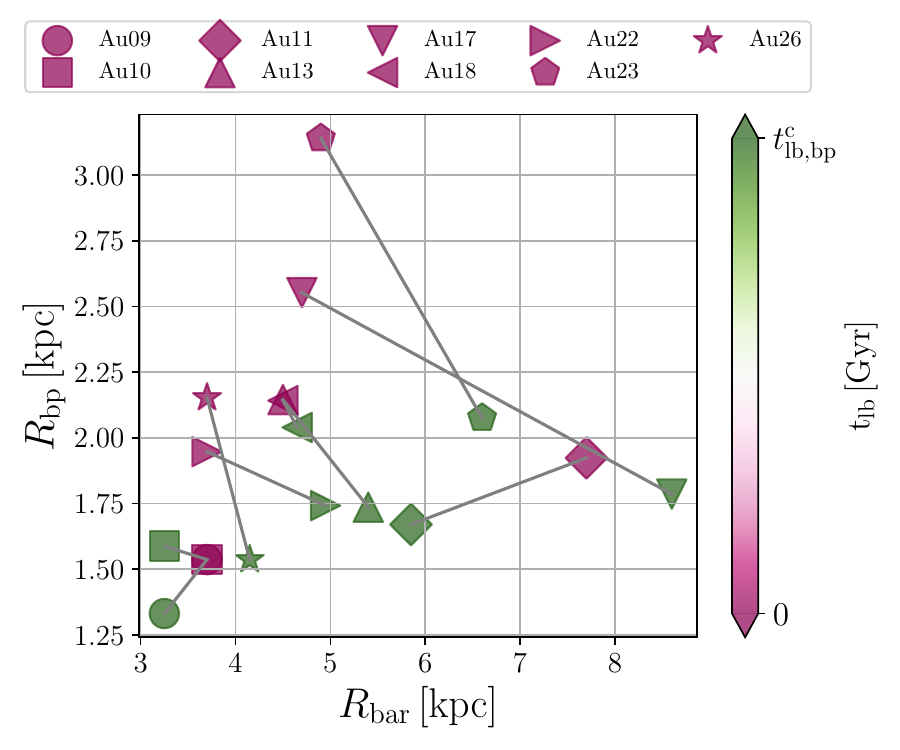}
    \caption{Scatter plot of the b/p size as a function of bar size for each galaxy, at two times: around the formation of the b/p bulge ($t_{\rm bp}^{\rm c}$) and at the end of the simulation ($t_{\rm lb} = 0 \, \rm Gry$).}
    \label{fig:rbp_rbar_scatter}
\end{figure}

To understand the behaviour of the $R_{\rm bp}/R_{\rm bar}$ ratio over time, we can refer to Fig. \ref{fig:rbp_rbar_scatter}. This figure shows the size of the b/p bulge as a function of the bar size for each galaxy in two moments: around the formation of the b/p structure ($t_{\rm bp}^{\rm c}$) and at the end of the simulation ($t_{\rm lb} = 0 \, \rm Gyr$). Observing the behaviour of most of the galaxies, we can distinguish a general trend: an increase in the size of the b/p, accompanied by a decrease in the bar size. Some \emph{N}-body studies have suggested that the bar length can decrease when the b/p's forms \citep{Martinez-Valpuesta_2004}, a phenomenon that we can see in these galaxies. There are some exceptions to this trend. On one hand, Au09 and Au11 share the same tendency of an increment in the size of the bar since the emergence of the b/p bulge. This seems to be related to the method used for measuring the bar size that sometimes can overestimate this quantity, meanwhile the b/p bulge of these galaxies is increasing. On the other hand, Au10 is having a decrease in the size of the b/p bulge since its formation, which is interesting because the b/p structure in this galaxy not only is becoming weaker but also it seems to be becoming shorter.

\begin{figure} 
    \centering
    \includegraphics[width=1\columnwidth]{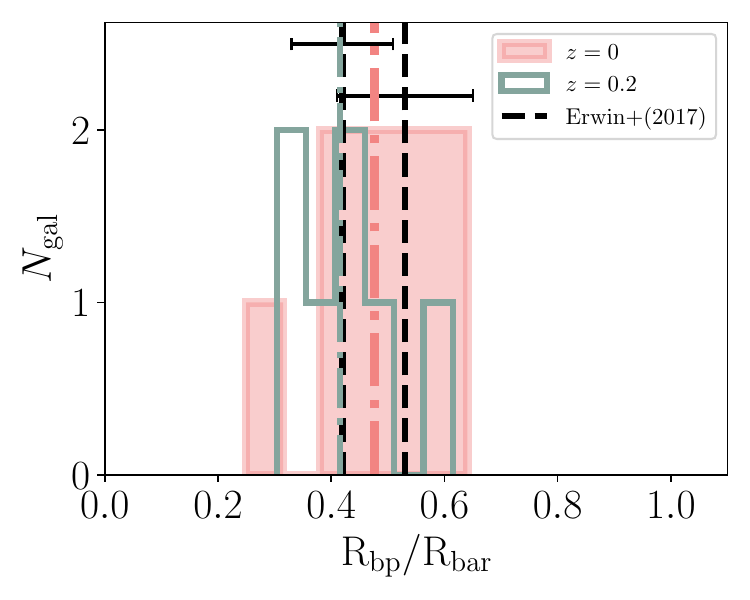}
    \caption{Histogram of the ratio between the b/p bulge radius and the bar radius at $z=0$ (pink) and $z=0.2$ (green). The dashed-dotted vertical lines indicate the median values for each redshift. The dashed vertical lines show the range of ratios measured by \citet{Erwin_2017}, with the right line representing the upper limit (bar size at maximum ellipticity) and the left line representing the lower limit (full bar extent), along with error bars.}
\label{fig:rbp_rbar_hist}
\end{figure}

Fig. \ref{fig:rbp_rbar_hist} presents a histogram of the $R_{\rm bp}/R_{\rm bar}$ values obtained at $z=0$ (pink) and at $z=0.2$ (green). 
The vertical dashed-dot lines represents the median for the two redshifts, finding $(R_{\rm bp}/R_{\rm bar})_{\rm median}^{z=0} = 0.48$ and $(R_{\rm bp}/R_{\rm bar})_{\rm median}^{z=0.2} = 0.41$.
This indicates that in this sample, the b/p bulge extends to nearly $50$ per cent of the bar's size at $z=0$. These results are consistent with those of the observations. We compare this result with the analysis performed by \cite{Erwin_2017}, where the authors examined $84$ barred galaxies from an initial sample of $186$ disc galaxies from the RC3 catalogue and the Virgo Cluster Catalogue \citep{Binggeli_1985}. They measured bar sizes using a combination of ellipse fitting and visual inspection, obtaining a lower limit $a_{\epsilon}$ and an upper limit $L_{\rm bar}$ for the bar size. Additionally, they provided an approximate measurement of the b/p bulge size $R_{\rm box}$, determined by the total visible extent of the main boxy region in logarithmic-scale images. With the size of the b/p and the two measurements for the bar size, they calculated the size ratio, finding $R_{\rm box}/a_{\epsilon} = 0.53 \pm 0.12$ (median = 0.54) and $R_{\rm box}/L_{\rm bar} = 0.42 \pm 0.09 $ (median = 0.43). These values are shown in Fig. \ref{fig:rbp_rbar_hist} as vertical dashed lines, along with the corresponding error bars. 
There is a very good agreement between the sizes in our sample and those found by the authors, even considering that the size measurement methods are different. 
On the other hand, \cite{Athanassoula_2015} analysed a sample of galaxies with barlens morphology (the face-on view of b/p bulges) from \cite{Laurikainen_2011}, estimating an average ratio between the sizes of the barlenses and the bars to be around $\sim 0.6$. Similarly, \cite{Lutticke_2000} measured the vertical thickness of b/p structures in a sample of galaxies and found a ratio of $\sim 0.4$. As a result, we also observe alignment between our findings and those obtained in numerical simulations.
Finally, we observe that by taking a larger redshift ($z=0.2$), the median shifts towards smaller values, which is consistent with the general trend that we observe of the bar length of the galaxies decrease after the b/p bulge is formed.

%-----------------------------------------------------------------
\section{Buckling episodes} \label{Sec:buckling_episodes}
Once bars form, they can trap more stars within them, growing longer in time and increasing in mass (see e.g. \citealt{Athanassoula_2002, Debattista_2006}). This increment can lead to a scenario that favours the emergence of instabilities. Small vertical perturbations can grow, causing the bar to bend, thus creating asymmetries around the disc plane. This process, known as buckling, results in the bar experiencing deformation and vertical asymmetry. After a buckling episode, the bar's morphology can be transformed from rather flat, to having an X-shaped or peanut-shaped structure, i.e. forming a b/p bulge. After the formation of bars, we aim to assess whether the galaxies experienced episodes of buckling during the formation process of the b/p structure or at a later time. 

\begin{figure*}
    \centering
    \includegraphics[width=2\columnwidth]{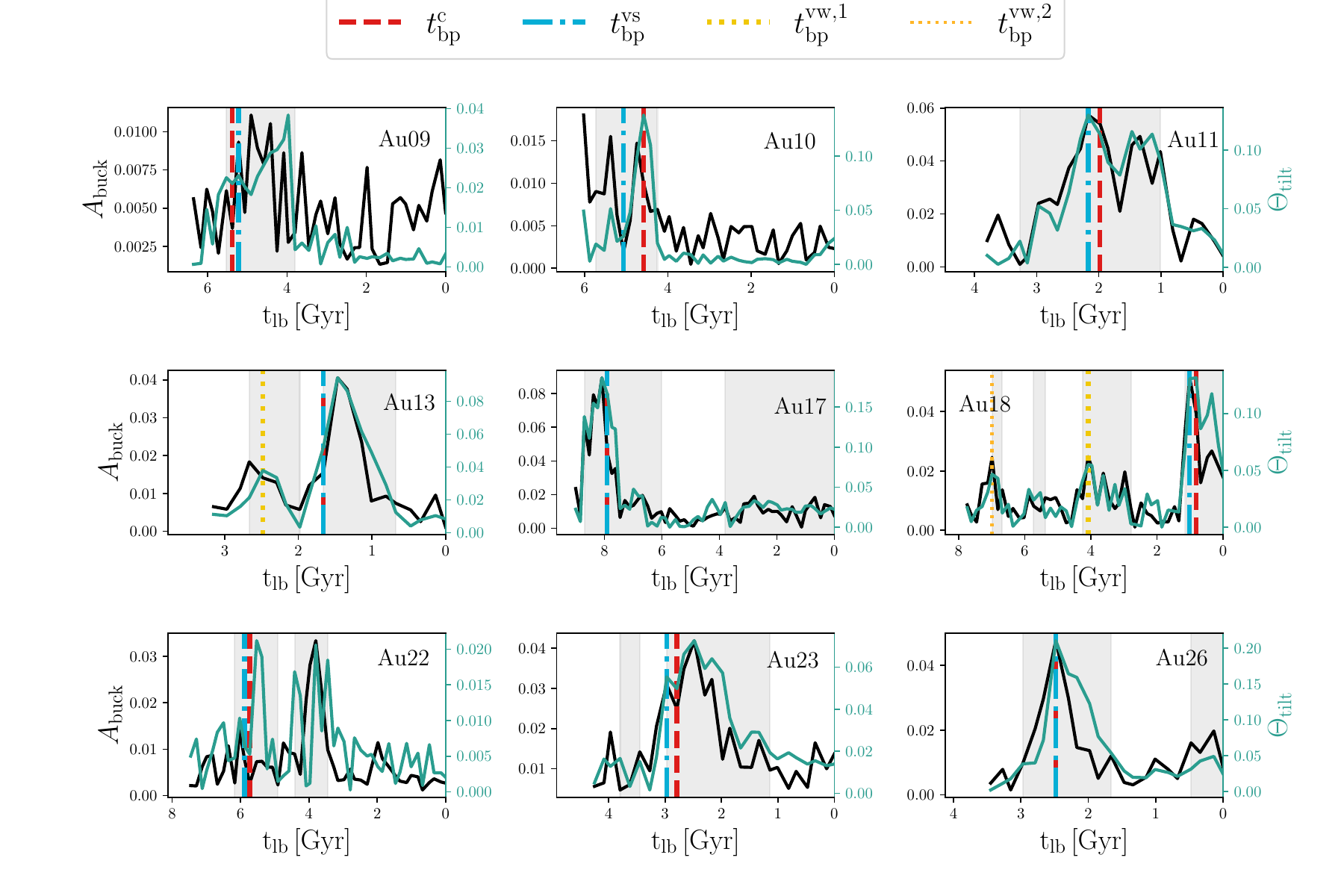}
    \caption{Evolution of the buckling amplitude, $A_{\rm buck}$, and the meridional tilt angle, $\Theta_{\rm tilt}$, for the galaxies in our sample. The buckling amplitude is represented by black lines with values marked on the left $y$-axis, while the tilt angle is shown by green lines with values indicated on the right $y$-axis. The evolutionary tracks start at the lookback time of bar formation, $t_{\rm{bar}}$, of each galaxy. The dashed line is the lookback time formation of the b/p bulge according to our method ($t_{\rm bp}^{\rm c}$), the dashed-dotted line is the lookback time formation of the visually strong b/p ($t_{\rm bp}^{\rm vs}$), the thin dotted line is the lookback time formation of the very visually weak b/p for Au18 ($t_{\rm bp}^{\rm vw,1}$), and the thick dotted line is the lookback time formation of the visually weak b/p ($t_{\rm bp}^{\rm vw,2}$). The shaded regions indicate the lookback times at which the bar is visually detected to be buckling.}
    \label{fig:buckling_all}
\end{figure*}

\begin{figure}
    \centering
    \includegraphics[width=1\columnwidth]{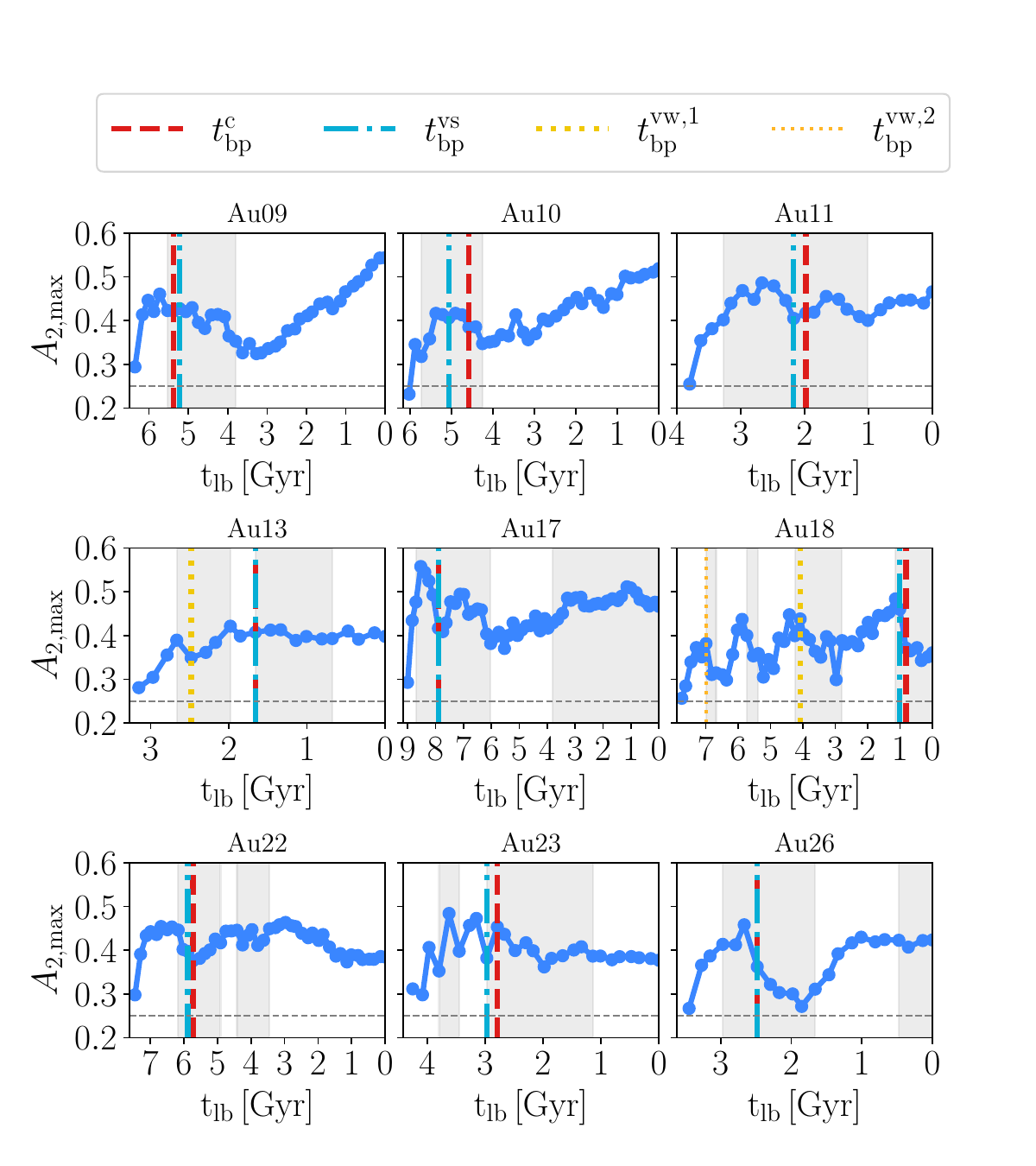}
    \caption{Bar strength as a function of lookback time from $t_{\rm{bar}}$ to $t_{\rm lb}=0 \, \rm Gyr$. The vertical lines are the different formations times for the b/p bulge: peak-detection method (dashed line, $t_{\rm bp}^{\rm c}$), visual strong (dashed-dotted line, $t_{\rm bp}^{\rm vs}$), visual weak (thick dotted line, $t_{\rm bp}^{\rm vw,1}$), visual very weak for Au18 (thin dotted line, $t_{\rm bp}^{\rm vw,2}$). The shaded regions indicate the lookback times at which the bar is visually detected to be buckling (see Sec. \ref{Sec:buckling_episodes}).}
    \label{fig:a2_buckling}
\end{figure}

To accomplish this, we identify the lookback times where an asymmetry relative to the major axis of the bar is observed (indicating buckling), by using different methods. The first method is a visual inspection of the edge-on projection of the stellar density and the unsharp mask, and the result is represented as the shaded regions in Figs. \ref{fig:bp_strength_time},  \ref{fig:bp_size}, \ref{fig:buckling_all}, and \ref{fig:a2_buckling}. We also employed a method that measures the buckling amplitude, labelled as  $A_{\rm buck}$ \citep{Sellwood_Athanassoula_1986,Debattista_2006,Debattista_2020}. This parameter quantifies the vertical buckling amplitude using the second harmonic of the azimuthal angle, weighted by the vertical positions and masses of the particles, and is defined as
\begin{equation}
    A_{\rm buck} = \left | \frac{\Sigma_j z_j m_j e^{2i\phi}}{\Sigma_j m_j} \right|,
\end{equation}
where $m_j$, $z_j$, and $\phi_j$ are the mass, vertical position, and azimuthal angle of the $j$th particle. The sum is over all the stars within the cuts, i.e. $|y| < 0.8 \, \rm kpc$, $|z| < 2 \, \rm kpc$, and $r < R_{\rm bar, 0}$. Finally, we also evaluated the meridional tilt angle, $\Theta_{\rm tilt}$ \citep{Ghosh_2024}. This parameter indicates the orientation of the velocity ellipsoid in the $R-z$ plane. A non-zero tilt angle suggests that the stellar velocities are correlated between the radial and vertical directions, implying a deformation of the velocity ellipsoid. The meridional tilt angle is defined as
\begin{equation}
    \Theta_{\rm tilt} = \frac{1}{2} \rm tan^{-1} \left ( \frac{2\sigma_{\rm{R}z}^2}{\sigma_{\rm RR}^2 - \sigma_{zz}^2}\right),
\end{equation}
where $\sigma_{ij}^2 = <v_iv_j>-<v_j><v_i>$ is the stellar velocity dispersion tensor \citep{Binney_Tremaine_2008}, using cylindrical coordinates over the stars within the cuts $|y| < 0.8$, $|z| < 2$, and $r < R_{\rm bar, 0}$.

\begin{figure*}
    \centering
    \includegraphics[width=2\columnwidth]{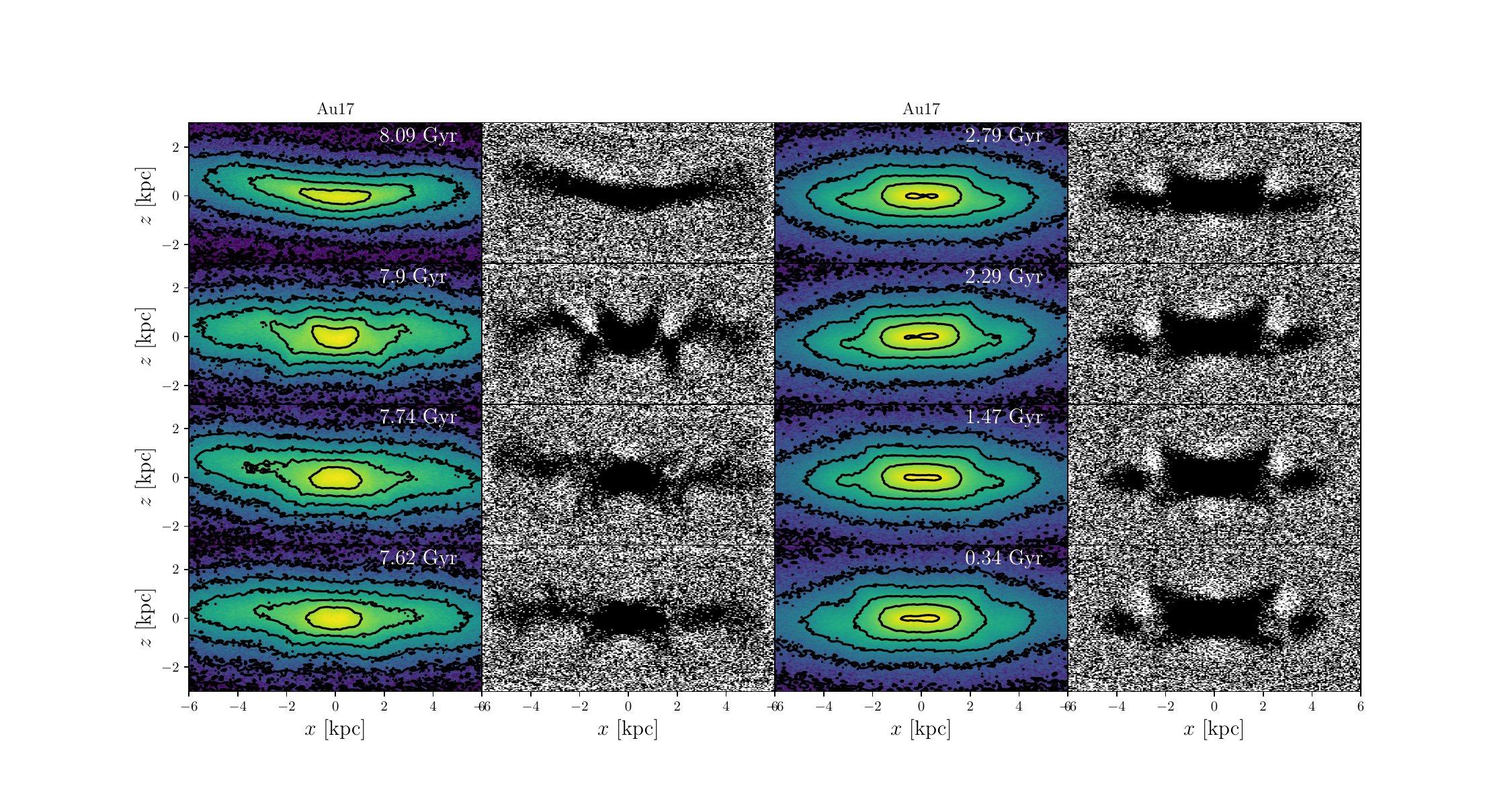}
    \caption{Stellar density distribution and unsharp mask in the edge-on projection, for two buckling episodes recorded for the simulated galaxy Au17. For the first buckling we display the interval going from $8.09-7.62 \, \rm Gyr$ in lookback time, and for the second buckling from $2.79-0.34 \, \rm Gyr$ in lookback time.}
    \label{fig:Au17_buckling}
\end{figure*}
Fig. \ref{fig:buckling_all} shows the temporal evolution of $A_{\rm buck}$  and $\Theta_{\rm tilt}$, with their values indicated on the left and right y-axes, respectively, starting from the formation time for the bar ($t_{\rm bar}$) for each galaxy. The vertical lines represent different definitions of the b/p formation times: the dashed line indicates the formation time of the b/p bulge as detected by the peak-detection method ($t_{\rm bp}^{\rm c}$), the dashed-dotted line shows the time of visually strong b/p bulge formation ($t_{\rm bp}^{\rm vs}$), the thick dotted line represents the visually weak b/p formation time ($t_{\rm bp}^{\rm vw,1}$), and the thin dotted line marks the formation time of a very weak b/p bulge ($t_{\rm bp}^{\rm vw,2}$), defined specifically for Au18. As mentioned before, the shaded regions represent the visually-identified periods of the buckling episodes.

In general, the evolution of both parameters exhibits similar trends, with peaks coinciding with the onset of visually identified buckling episodes. 
$A_{\rm buck}$ seems to be a more noisy parameter. For example in Au09 and Au10 it shows peaks where there is not a visually-identified buckling. On the other hand, if we look at Au22, $A_{\rm buck}$ is not detecting the first buckling episode recorded. Meanwhile, $\Theta_{\rm tilt}$ has a smoother evolution, showing clear peaks in the time periods where we identified visually the presence of a  buckling. This suggests that, while both parameters generally behave similarly, $\Theta_{\rm tilt}$ is more effective at capturing buckling events.

We find that all galaxies experienced buckling episodes before or around the formation of the b/p bulge, and we did not find any case where the b/p structure formed without a break in mid-plane symmetry in the edge-on projection. This strongly suggests that the formation mechanism of these structures in these simulations is intimately linked to the buckling instability.
Furthermore, more than half of the galaxies exhibiting double buckling events and in particular one of them present a quadruple buckling episode (Au18). Among these, Au17, Au18, and Au26 are undergoing buckling at $z=0$. \cite{Martinez-Valpuesta_2006} had previously identified the possibility of bars experiencing double buckling episodes, where bars undergoing a vertical buckling instability initially weaken, followed by renewed growth leading to recurrent buckling. In the case of Au18, the behaviour remains unique. As previously mentioned, a structure resembling a b/p is observed around $t_{\rm lb} \sim 7 \, \rm Gyr$, accompanied by an asymmetry in the edge-on projection. Another episode is recorded around $t_{\rm lb} \sim 5.7 \, \rm Gyr$ and another around $t_{\rm lb} \sim 4.2 \, \rm Gyr$ close to the visually weak b/p formation time, with a final episode occurring in the last gigayear of its evolution. In total, this galaxy experiences four buckling episodes, but the first two are very weak and last in between $0.18 - 0.35 \, \rm Gyr$. Interestingly, during these visually recorded episodes, peaks are observed in the $A_{\rm buck}$ and $\Theta_{\rm tilt}$ parameters, though their amplitudes are much smaller for the first two episodes. The parameters generally show a good agreement with the visually identified episodes, although they do not have a signal for all visually identified episodes. 

Another characteristic to consider regarding these episodes is the timescale over which they occur, showing significant diversity. The shortest episode lasts $0.31\,{\rm Gyr}$ for Au22, while Au17, experiencing double buckling episodes, exhibits the longest duration: one lasting $\sim 2.4\,{\rm Gyr}$  and the other $3.8\,{\rm Gyr}$. Fig. \ref{fig:Au17_buckling} illustrates a part of the two buckling episodes experienced by the bar of Au17: the first buckling goes from $8.7-6 \, \rm Gyr$ and the second goes from $3.8-0 \, \rm Gyr$. It is interesting to note that the time interval between the first episode and the second episode is $2.2\,{\rm Gyr}$, roughly matching the duration of the buckling. The bar's edge-on morphology changes considerably during the lifetime of this galaxy. The entire disc is deformed during the first buckling episode, until a strong b/p is formed. Afterwards, the b/p strength diminishes significantly (see also Fig. \ref{fig:bp_strength_time}) to the point that the b/p is no longer detectable, not even with visual inspection, i.e. the b/p is in a sense, destroyed, or attenuated. After that, at around $t_{\rm lb} \sim 3.8\,{\rm Gyr}$, the bar experiences a second buckling episode, which makes the bar asymmetric, and which is ongoing at $z=0$.

Fig. \ref{fig:a2_buckling} shows how the bar strength of the galaxies was affected by the buckling episodes and the formation of the b/p bulges. The buckling starts at $t_{\rm lb}$ when the bar is strong and well-defined, in agreement with works that use \emph{N}-body simulations \citep{Debattista_2020,Li_2023}, although it does not necessarily always coincide with the maximum value of its strength. For galaxies experiencing double buckling episodes, we observe different behaviours. In Au18, the bar weakens each time buckling occurs, but gains strength in between episodes. In contrast, the second buckling episode in Au17, Au22, and Au26 does not seem to affect the bar as much as the first, which caused significant weakening. These galaxies begin to regain strength as soon as the first episode ends. 
The second episode, while not affecting Au13, causes Au23's bar to weaken until it stabilizes halfway through the event, maintaining a similar strength until $z=0$. For galaxies with a single buckling episode, Au09 and Au10 exhibit a similar trend: the bar weakens during the buckling but steadily gains strength afterward, reaching a maximum at $z=0$. Au11, however, shows an increase in bar strength during buckling, followed by a slight weakening when the b/p forms, and then regains strength after the episode. We find that at times, the bar strength can increase at the onset of the buckling episodes, but is weakened once the b/p bulge forms.

\section{Discussion} \label{Sec:Discussion}
%-----------------------------------------------------------------------------------------

\subsection{Delay time between bar and b/p bulge formation}

When using our automatic b/p bulge detection method, we find a median delay time between the formation of the bar and the formation of the b/p structure of $1.6~\rm{Gyr}$. When considering the first appearance of the b/p bulge -- using either visual inspection or our automated method -- we find a median delay time of $1.1~\rm{Gyr}$. Therefore, b/p structures in the Auriga simulations typically form between $1.1-1.6~\rm{Gyr}$ after the formation of the bar (Fig. \ref{fig:delta_t}). This delay time is consistent with observational results, such as from \cite{Kruk_2019}; in their work, they compare high-redshift observations from \cite{Simmons_2014} that identify barred galaxies at $z = 1.5-2$, suggesting these bars formed at around $\sim 9-10 \, \rm Gyr$ ago. Based on observations of b/p bulges at higher redshifts ($z \sim 0.7-0.8$), they estimate a b/p formation time around $\sim 7\, \rm Gyr$ ago. They argue that this therefore indicates a delay of approximately $2\, \rm Gyr$ between the formation of the earliest bar and b/p structures. However, caution is needed in this interpretation, since recent studies have found barred galaxies at even higher redshifts, up to $z \sim 3$ or even $\sim 4$ (e.g. \citealt{Costantin_2023,Guo_2024, Le_Conte_2024}), which corresponds to a lookback time of $\sim 12\, \rm Gyr$, which would suggest an even earlier formation time for bars. Additionally, it is not trivial to draw a direct causal relation between the first barred galaxies and the first b/p's; for example, some early forming bars might dissolve (e.g. \citealt{Kraljic_2012}) before forming a b/p bulge. The oldest b/p bulge in the Auriga simulations forms at a lookback time of $\sim 8~\rm{Gyr}$, while the majority of the b/p structures formed between $2.5-5.8~\rm{Gyr}$ ago, with a couple of galaxies forming the structure within the last $2~\rm Gyr$ (median = $3.8~\rm Gyr$; see Fig. \ref{fig:tvs_tcode}). These results are consistent with the aforementioned study of \cite{Kruk_2019}, who find  b/p bulges in galaxies at lookback times $3-8 \, \rm Gyr$.

We now turn our attention to comparing our results with previous studies of numerical simulations.  \cite{Combes_1990} use isolated \emph{N}-body simulations to study the formation of b/p structures, and find a delay between the formation of the bar and the b/p bulge of $\sim 2 \, \rm{Gyr}$. \cite{Pfenniger_1991} analyse in detail a \emph{N}-body barred galaxy and find that the bar becomes peanut-shaped at approximately $1.68 - 1.78 \, \rm{Gyr}$ after its formation, while \cite{Athanassoula_2008} show that a b/p bulge forms about $\sim 1 \, \rm{Gyr}$ after the bar. These studies are all roughly in line with the delay time we find in the Auriga simulations. Furthermore, \cite{Ghosh_2024} consider isolated N-body galaxy models, with varying thin and thick disc proportions, and find that there is a shorter delay between bar and b/p bulge formation for an increasing mass fraction of the thick disc (i.e. galaxies with a more massive thick disc form b/p's more rapidly); depending on the model, the time difference between bar and b/p formation ranges between $0.5 - 3 \, \rm{Gyr}$.

\subsection{Multiple buckling events}

We find that all the Auriga galaxies experience a break in symmetry along the x-z plane prior to the emergence of the b/p bulge. This suggests that the vertical asymmetric buckling of bars is an important mechanism for the formation of b/p's in these simulations. From a sample of $84$ local barred galaxies, \cite{Erwin_2016} identify two galaxies with morphological and kinematic evidence that are currently undergoing a buckling episode. These galaxies belong to a subsample of $44$ high-mass galaxies ($\rm{log}(M_{\rm{star}}) \gtrsim 10.4 \, M_{\odot}$, which is the same mass range as the galaxies in our sample), thus obtaining an observed fraction of local barred galaxies undergoing a buckling event of $4.5$ per cent. In the Auriga simulations, we find that $3$ of the $9$ galaxies are experiencing a buckling episode at $z=0$, i.e. $1/3$ of the galaxies, a significantly higher fraction than in the aforementioned observations. In the same study, the authors also developed a simple model of galaxy evolution to study the buckling phase, obtaining a time difference between bar formation and the first buckling of approximately $\sim 1-2~\rm{Gyr}$, while the buckling episode had a duration of $\sim 0.5-1~\rm{Gyr}$. In the case of galaxies in the Auriga simulations, we find a delay between bar formation and buckling of $\sim 0.4-1.5~\rm{Gyr}$ with buckling episodes lasting between $\sim 0.3-3.8~\rm{Gyr}$. These results are also consistent with the ones found for isolated {\em N}-body simulations in the work of \cite{Martinez-Valpuesta_2006}, where the authors find recurrent buckling events with a duration of $\sim 1$ and $\sim 3 \, \rm Gyr$ each, with a time difference between the two events of $\sim 4.4$, twice as long as the time between Au17 bucklings ($2.2 \, \rm Gyr$).

B/p bulges do not always form via a buckling instability, as they can also form via gradual mechanisms such as resonant trapping of stars (e.g. \citealt{Quillen_2002}). However, as previously noted, in the Auriga simulations we find that all the b/p structures are formed after a buckling event. To understand why this is happening, we turn to previous works that explore under which circumstances gradual resonant trapping can occur. \cite{Sellwood_2020}, using \emph{N}-body simulations, found that the presence of a nuclear star cluster (a dense central mass) can inhibit the occurrence of buckling and causes stars to gradually puff up out of the plane. Some studies use simulations that include a gas component and they report that gas has an effect on buckling. For example, \cite{Baba_2022} uses an \emph{N}-body/hydrodynamics simulation of an isolated galactic disc that includes gas, radiative cooling, and star formation, and finds that the inclusion of both a gas component and star formation results in the formation of a b/p bulge without buckling. In addition, other works \citep{Berentzen_1998, Debattista_2006} agree that gas can significantly influence the evolution of discs, in particular by generating an increase in the central mass concentration (CMC) which, as \cite{Sellwood_2020} found, results in the suppression of buckling. We therefore see that a massive CMC can contribute to the suppression of the buckling instability in the disc. The reason that we do not see this gradual heating occurring in the Auriga simulations, could be due to the lack of massive CMCs in the centres of galaxies. For example, the interstellar medium (ISM) model used in the simulations \citep{Springel_2003}, together with the star formation and feedback model, might not allow a very dense central mass concentration to form. We defer a detailed exploration of the properties of CMCs in Auriga and their effect on disc stability to future work.

\subsection{b/p fraction and comparison with observations}

Finally, we discuss our findings in terms of the b/p fraction over cosmic history. At $z=0$, we find that $45$ per cent of barred galaxies host a b/p bulge. This value lies between the range of observational estimates (see Fig. \ref{fig:bp_fraction}), although the observed fractions vary significantly between studies; this suggests that observational limitations and methodological differences play a critical role in  determining the b/p fraction. For example, \cite{Erwin_2017}, who report the highest b/p fraction out of the studies we consider, find that $80$ per cent of massive barred galaxies host a b/p bulge at $z=0$. On the other hand, \cite{Kruk_2019} quote two b/p fractions in their study: a detected b/p fraction, which is $31$ per cent, and a corrected fraction of $69$ per cent, which aims to account for biases such as image resolution, signal-to-noise limitations, and band-shifting effects that hinder the detection of b/p structures at high redshift. They do this by simulating how nearby SDSS galaxies would appear at greater distances, and then estimate how many b/p structures are missed, adjusting their fraction accordingly. Other studies show similarly wide variations. \cite{Lutticke_2000} report a fraction of $68$ per cent at $z=0$, closely matching the corrected fraction of \cite{Kruk_2019}, whereas \cite{Yoshino_Yamauchi_2015} find a significantly lower fraction of $33$ per cent. This fractions are multiplied by a factor of 3/2, to take into account that approximately two-thirds of disc galaxies in the local Universe host bars.
\cite{Yoshino_Yamauchi_2015} discuss the discrepancies between their results and those of \cite{Lutticke_2000}, suggesting that the variations may be attributed to several factors: the inherent difficulty of detecting b/p structures, the quality of observational data, and different definitions of what constitutes a b/p bulge. They note that the sample of \cite{Lutticke_2000} mainly consists of late-type galaxies, where the bulges appear ambiguous in the Digitized Sky Survey (DSS) images. Furthermore, many late-type galaxies have their bulges classified as `\textit{close to box-shaped, not elliptical}’, leading the authors to suggest that the b/p fraction in the work of \cite{Lutticke_2000} may be overestimated.

The results of \cite{Erwin_2017} (who find a b/p fraction of 80\%), point to a higher b/p fraction than the one we find in Auriga (45\%), for galaxies in the same stellar mass regime. Putting aside the observational uncertainties in determining b/p fractions, this might indicate that the Auriga simulations are under-producing b/p bulges.
As discussed in \ref{Sec:bp_fraction}, the ratio of vertical to radial stellar velocity dispersion ($\sigma_z / \sigma_R$) plays a crucial role in determining the stability of the disc against buckling instability (e.g. \citealt{Merritt_1994,Martinez-Valpuesta_2006}), with the formation of b/p bulges favoured in galaxies with low $\sigma_z / \sigma_R$. A low ratio implies that the stars have a greater radial motion compared to the vertical motion, producing a centrifugal force that dominates over the gravitational force of the disc, which is trying to confine the stars to the plane. For example, in the study of \cite{Ghosh_2024} they showed -- using isolated {\em N}-body simulations --that galaxies with hotter discs tend to form weaker b/p's. In the Auriga simulations, the discs tend to have slightly high $\sigma_z$ (see e.g. \citealt{Grand_2016}), which therefore could make it harder for buckling to occur, or could produce b/p's that are weaker. There are several mechanisms that might be driving the vertical heating of discs in Auriga, such as external perturbations of satellite galaxies, while the effective equation of state used for the ISM \citep{Springel_2003} could also be playing an important role in limiting how thin the disc can be. Although we show in Sec. \ref{Sec:bp_fraction} a general trend that barred galaxies without a b/p bulge have higher $\sigma_z / \sigma_R$, we defer a detailed exploration in the role of disc heating in b/p formation within the cosmological context to future work.

%-----------------------------------------------------------------
\section{Summary and conclusions} \label{Sec:Summ&Conclu}
%-----------------------------------------------------------------

In this work, we use the Auriga suite of cosmological zoom-in simulations to study the formation and evolution of boxy/peanut (b/p) bulges over cosmic history. To identify these structures, we use visual inspection and develop an algorithm to automatically identify the presence and strength of a b/p bulge. Additionally, we explore the properties of the b/p bulges, and how they relate to the properties of bars. 
Our results can be summarised as follows.
\begin{itemize}
    \item \textbf{b/p fraction:} We find that $45$ per cent of barred galaxies in Auriga at $z=0$ have a b/p bulge. This fraction decreases to $20$ per cent at $z=0.5$, and by $z=1$, none of the barred galaxies host a b/p bulge (Fig. \ref{fig:bp_fraction}).
    
    \item \textbf{b/p formation time:} 
    We find a variety of formation times for the b/p bulges; the oldest b/p forms at a lookback time of $t_{\rm lb} \sim 8 \, \rm Gyr$, with a majority of b/p's forming between $t_{\rm lb} \sim  2.5-5.8 \, \rm Gyr$. Two of the b/p's form within the last $2 \, \rm Gyr$.
    We find that the b/p bulges typically form between $1.1-1.6 \, \rm Gyr$ after the formation of the bar (Fig. \ref{fig:delta_t}).

    \item \textbf{Formation mechanism:} In all the galaxies hosting a b/p bulge, a buckling episode precedes the formation of the b/p structure, suggesting that this mechanism drives the formation of b/p's in cosmological simulations (see e.g. Figs. \ref{fig:bp_strength_time} \& \ref{fig:buckling_all}). Furthermore, galaxies that form b/p's tend to have older bars and a lower $\sigma_z/\sigma_r$ ratio, as compared to barred galaxies that do not form a b/p bulge (Fig. \ref{fig:sigmazr}).

    \item \textbf{b/p re-emergence:} We find two galaxies in which the b/p forms, then appears to dissolve (or is undetectable) and then reforms following a later buckling event. This suggests that these structures might emerge, dissolve, and re-emerge in a cosmological setting (see e.g. Fig. \ref{fig:bp_strength_time}).

    \item \textbf{Sizes of b/p bulges and bars:} A correlation appears to exist between the growth of the b/p bulge and a relative decrease in the size of the bar, consistent with their joint evolution (Fig. \ref{fig:bp_size} and Fig. \ref{fig:rbp_rbar_scatter}). By measuring $R_{\rm bp}/R_{\rm bar}$ at $z=0$, we found that the sizes of the b/p bulges represent about $\sim 50$ per cent of the size of their bars (Fig. \ref{fig:rbp_rbar_hist}).

    \item \textbf{Multiple buckling events:} More than half of the galaxies have double buckling episodes, with one of them experiencing four of these events (two of them happening during short periods of time) (Fig. \ref{fig:buckling_all} and Fig. \ref{fig:Au17_buckling}).
    
    \item \textbf{Impact of buckling on bars:} Galaxies that experience multiple or single buckling episodes show varying responses in their bar strength. We see that at the onset of the asymmetry, the bar can keep growing stronger, but once the b/p bulge forms, it weakens. Interestingly we find that, while some galaxies experience buckling events during brief periods of time, the majority of the events seems to remain during long periods of time (Fig. \ref{fig:a2_buckling}).   
    
\end{itemize}

With this study we explore in detail the formation and evolution of b/p bulges in the Auriga cosmological simulations. These structures are intimately linked to the evolution of bars and reveal important information about the dynamical state of the disc in the cosmological context. The fact that we find an underestimation of b/p's in the Auriga simulations compared to some observations hints to aspects of the galaxy formation and evolution model that stabilise the bar against forming these structures, such as the disc thickness. With a general view of these structures now established, future work will focus on a more detailed study of how b/p bulges form and influence stellar populations that emerge after they are in place.

\section*{Acknowledgements}

We thank the referee for their constructive report and comments. We thank Marie Martig for interesting discussions. PDL has received financial support from the European Union's HORIZON-MSCA-2021-SE-01 Research and Innovation programme under the Marie Sklodowska-Curie grant agreement number 101086388 - Project acronym: LACEGAL. This work used the DiRAC@Durham facility managed by the Institute for Computational Cosmology on behalf of the STFC DiRAC HPC Facility (www.dirac.ac.uk). The equipment was funded by BEIS capital funding via STFC capital grants ST/K00042X/1, ST/P002293/1, ST/R002371/1 and ST/S002502/1, Durham University and STFC operations grant ST/R000832/1. DiRAC is part of the National e-Infrastructure. FF is supported by a UKRI Future Leaders Fellowship (grant no. MR/X033740/1). This work was supported by STFC [ST/X001075/1]. CS and SAC acknowledge funding from Consejo Nacional de Investigaciones Científicas y Tecnológicas (CONICET, PIP KE4-11220200102876CO and PIP KE3-11220210100595CO), and Agencia Nacional de Promoción de la Investigación, el Desarrollo Tecnológico y la Innovación (Agencia I+D+i, PICT-2021-GRF-TI-00290). SAC acknowledges support from the Universidad Nacional de La Plata (G11-183), Argentina. RG is supported by an STFC Ernest Rutherford Fellowship (ST/W003643/1). FAG acknowledges support from ANID CYT Regular 1211370, the ANID BASAL project FB210003 and the HORIZON-MSCA-2021-SE-01 Research and Innovation Programme under the Marie Sklodowska-Curie grant agreement number 101086388. 

%%%%%%%%%%%%%%%%%%%%%%%%%%%%%%%%%%%%%%%%%%%%%%%%%%
\section*{Data Availability}

The data utilized in this study are publicly accessible for download at \href{https://wwwmpa.mpa-garching.mpg.de/auriga/data.html}{https://wwwmpa.mpa-garching.mpg.de/auriga/data.html}.



\bibliographystyle{mnras}
\bibliography{references} % if your bibtex file is called example.bib

%%%%%%%%%%%%%%%%% APPENDICES %%%%%%%%%%%%%%%%%%%%%

\appendix
\section{Sample of B/P bulges at z=0}
\label{App:bp_z0}

We show the stellar density distribution (and unsharp mask) in the edge-on projection for all the galaxies at $z=0$. Additionally, the radial profile of $\varcal{Z}$ with the location of the maximum, i.e. $R_{\rm bp}$, calculated using the method described in Sec. \ref{Sec:Method}.

\begin{figure*}
    \centering
    \includegraphics[width=1.5\columnwidth]{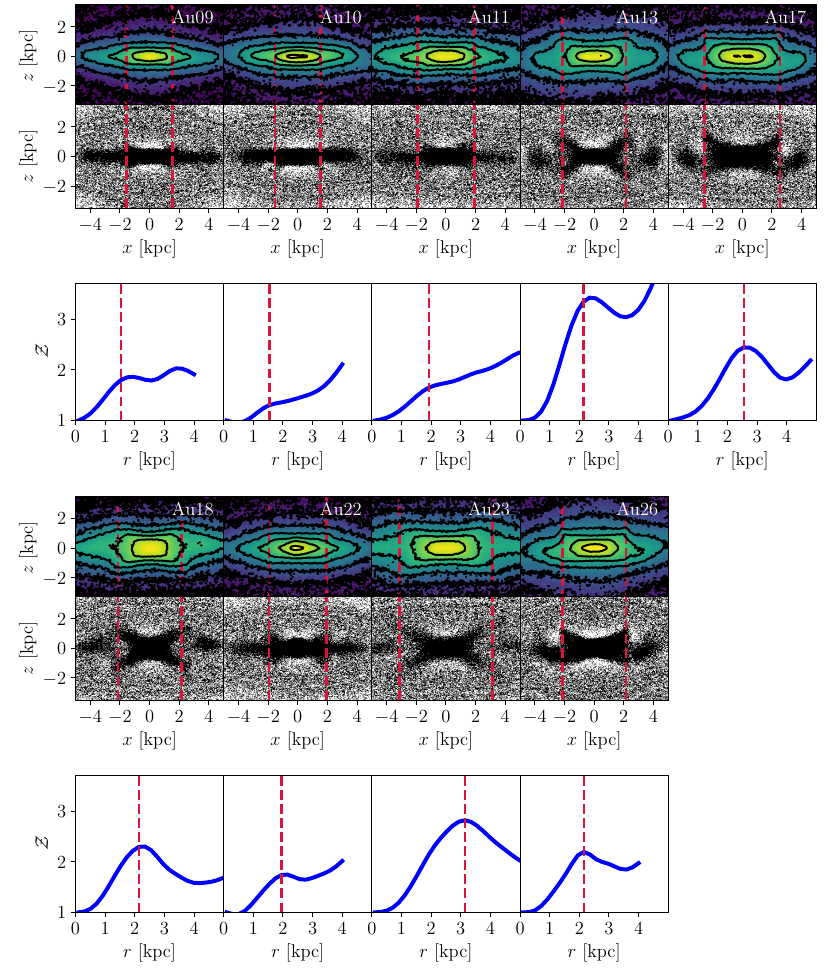}
    \caption{\textit{Upper panels}: Stellar surface density in the edge-on projection ($xz$ plane) for the galaxies of our sample, at $z=0$. \textit{Middle panels}: Unsharp mask of the surface density in the edge-on projection. \textit{Lower panels}: Radial profile of the median of the absolute value of the heights of the stars, normalised by the central value, at $z=0$. The vertical dashed lines in all the panels represent the size of the b/p bulges.}
    \label{fig:density_unsharp_plots}
\end{figure*}

\section{B/P bulge detection}
\label{App:bp_detection}
The different panels of Fig. \ref{App:bp_detection} shows how the minimum value of the second derivative of the parameter $\varcal{Z}$ (defined as the median of the absolute values of the $z$ positions of the stars that are part of the bar, normalized with the corresponding central value), $(\mathrm{d} ^2 \varcal{Z}(r)/\mathrm{d}r^2)_{\mathrm {min}}$, varies with lookback time for each galaxy. This value directly measures the curvature of $\varcal{Z}$: when negative, it indicates downward convexity corresponding to the presence of a maximum point. The more negative the value, the greater the curvature, resulting in a more pronounced peak. We observe that there is a moment when this value becomes significantly more negative and remains so over several snapshots (at least 5). Specifically, we find that a threshold of $(\mathrm{d} ^2 \varcal{Z}(r)/\mathrm{d}r^2)_{\mathrm {min}} = -0.51$ marks this drop, and we use it to define the lookback time where the b/p emerge: $t_{\rm bp}^{\rm c}$.
\begin{figure*}
    \centering
    \includegraphics[width=2\columnwidth]{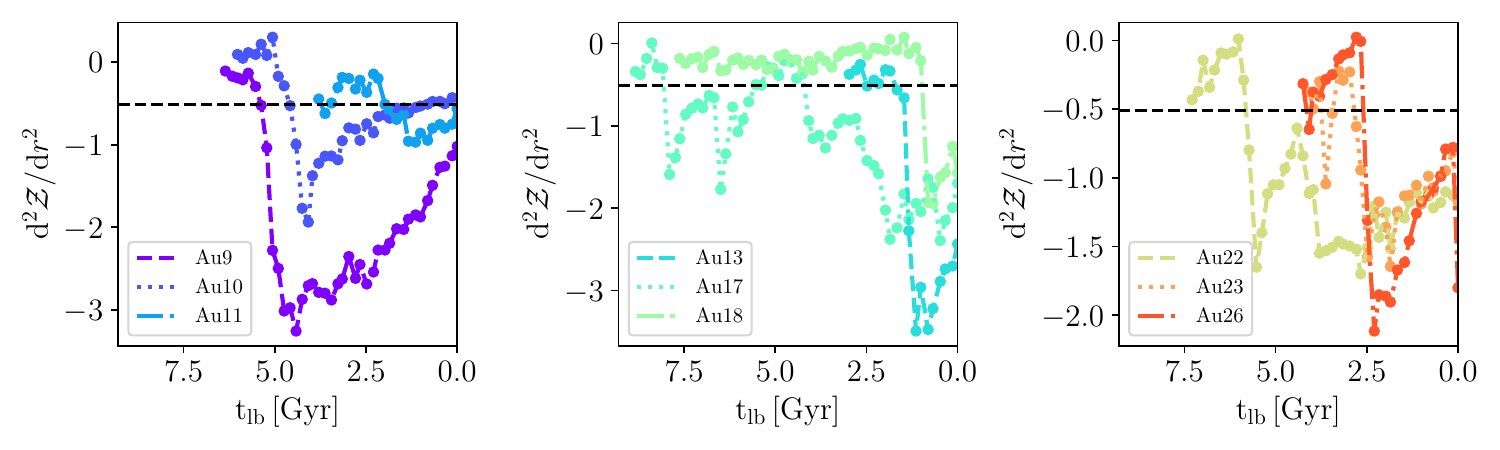}
    \caption{Minimum of the second derivative of the parameter $\varcal{Z}$, defined as the median of the absolute values of the z positions of the stars that are part of the bar, as a function of lookback time, for all the galaxies in our sample.}
    \label{fig:nabla_2_min}
\end{figure*}

\section{Weak b/p bulge in Au18}
\label{App:bp_weak}

We show the stellar density distribution (and unsharp mask) in the edge-on projection for Au18 for the times when we observe a weak b/p bulge, at $t_{\rm lb} \sim 4 \, \rm Gyr$ and $t_{\rm lb} \sim 7 \, \rm Gyr$.

\begin{figure*}
    \centering
    \includegraphics[width=2\columnwidth]{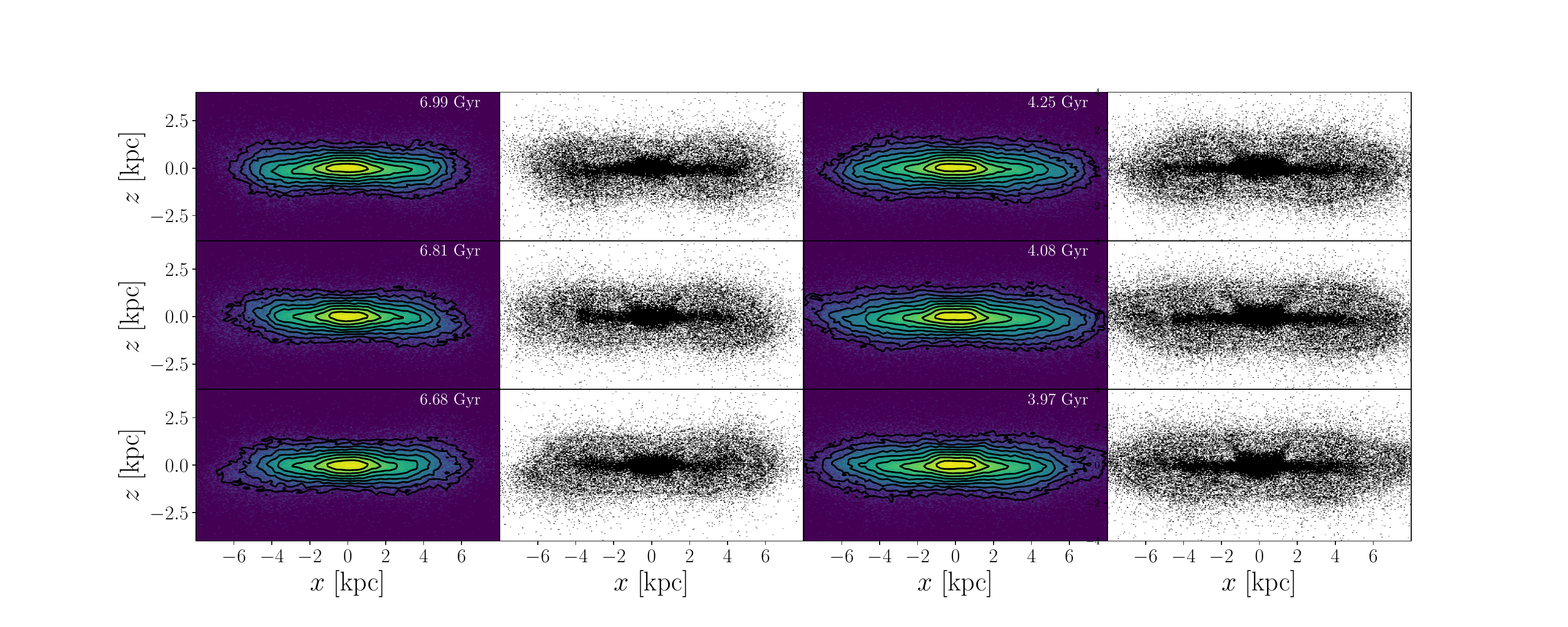}
    \caption{Stellar surface density and unsharp mask for Au18, during the periods when a weak b/p bulge is visually detected. The plots are shown with a cut-off of $|y| < 1 \, \rm kpc$ for better visualisation.}
    \label{fig:Au18_bpWeak}
\end{figure*}

\section{B/P fading}
\label{App:bp_fading}
\begin{figure*}
    \centering
    \includegraphics[width=2\columnwidth]{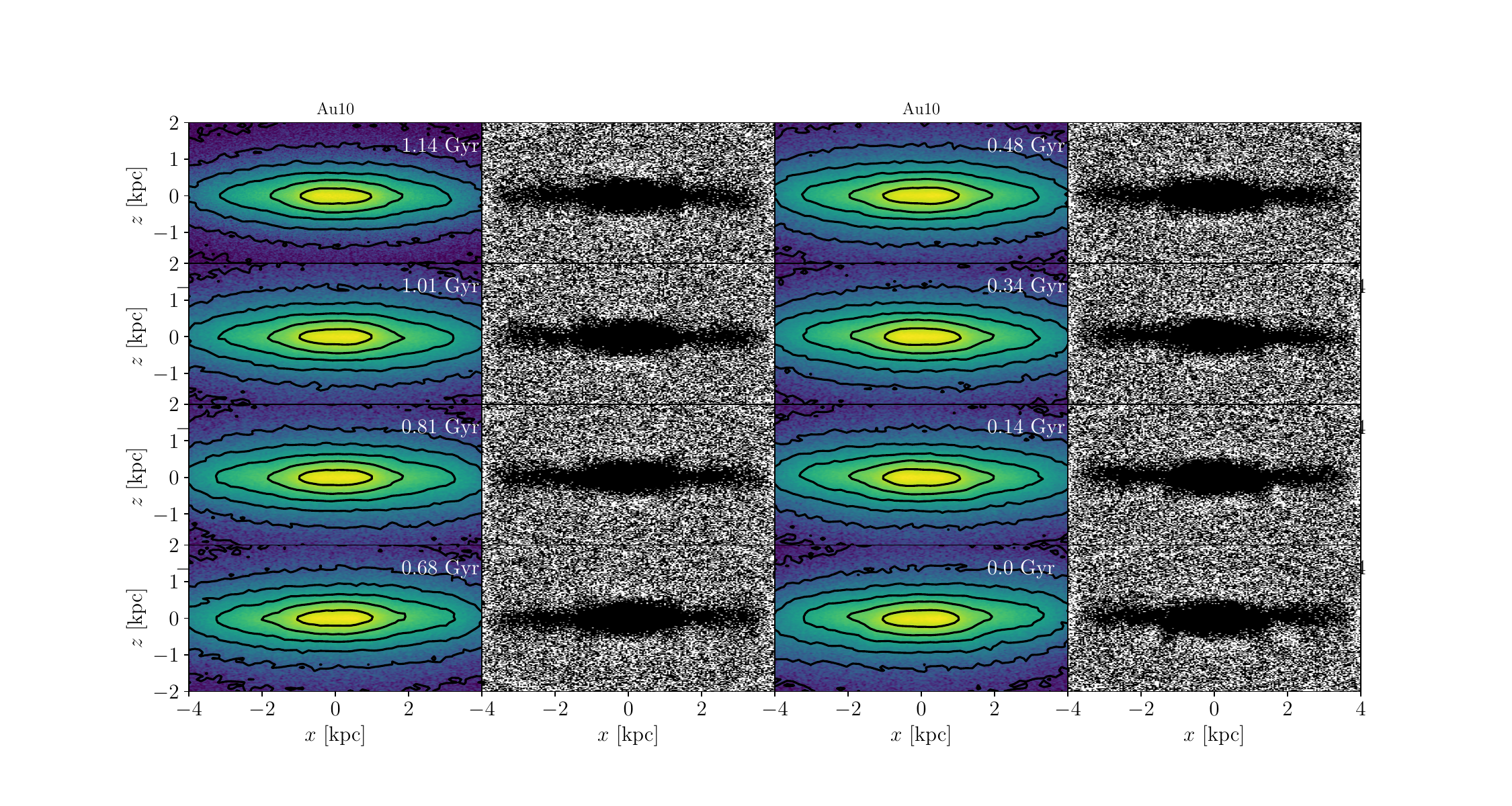}
    \caption{Stellar surface density and unsharp mask for Au10, during the periods when no strength associated with the b/p was detected: from lookback time $\sim 1.14-0.14 \, \rm Gyr$.}
    \label{fig:Au10_desapears}
\end{figure*}

\begin{figure*}
    \centering
    \includegraphics[width=2\columnwidth]{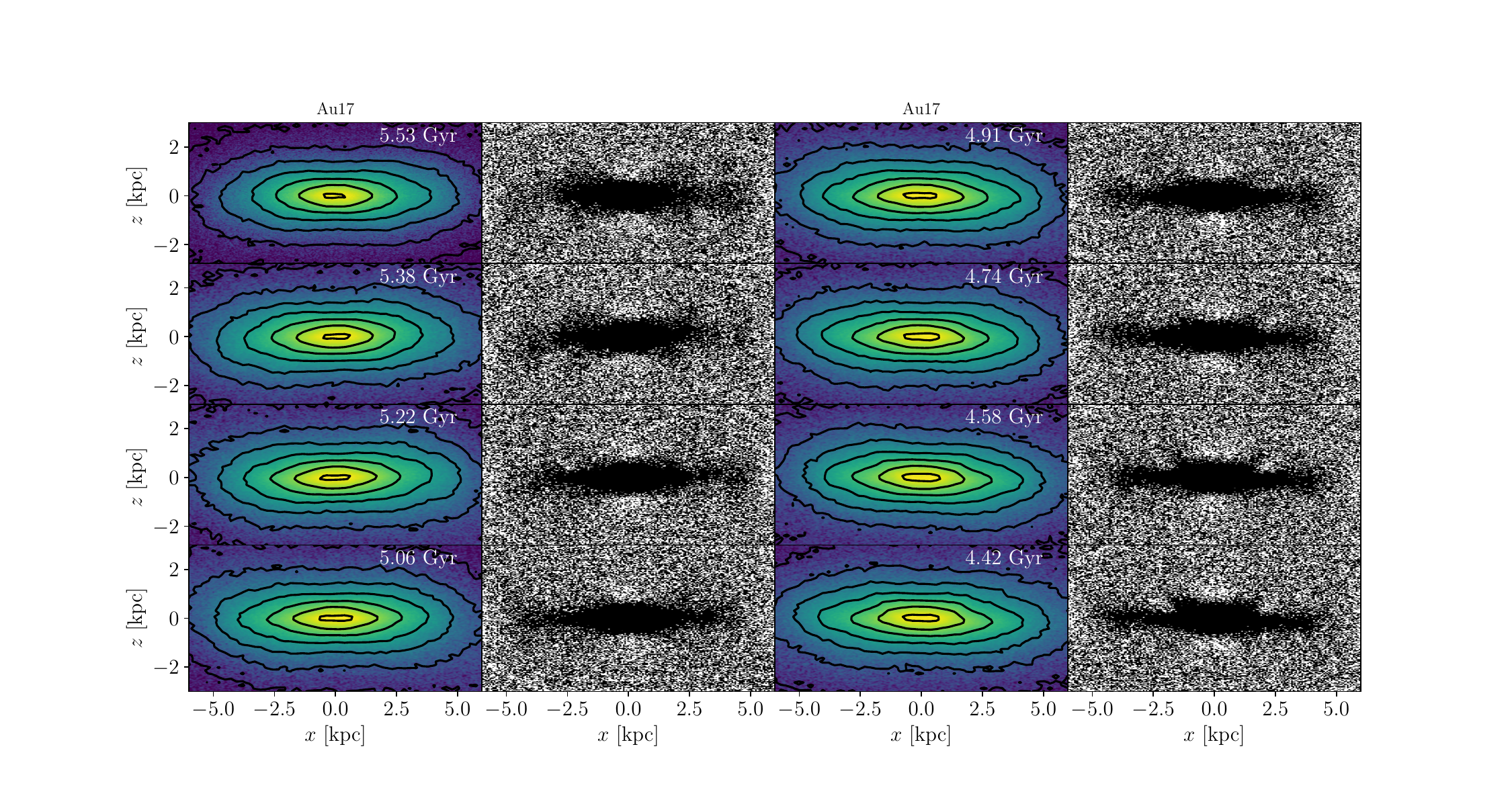}
    \caption{Stellar surface density and unsharp mask for Au17, during the periods when no strength associated with the b/p was detected: from lookback time $\sim 5.53-4.58 \, \rm Gyr$.}
    \label{fig:Au17_desapears}
\end{figure*}

Both Au10 and Au17 present moments where the strength associated with the b/p is zero. This can be understood by observing Fig. \ref{fig:nabla_2_min}, where after identifying the b/p formation lookback time, there are periods when the minimum of $(\mathrm{d} ^2 \varcal{Z}(r)/\mathrm{d}r^2)_{\mathrm{min}}$ takes values above $-0.51$. Fig. \ref{fig:Au10_desapears} and Fig. \ref{fig:Au17_desapears} show the stellar density distribution and the unsharp mask corresponding to each galaxy during the times when no associated strength was detected. These figures can help us understand what is happening. Initially, this could be due to two possibilities: either the structure weakens over time and the method fails to capture it, or the structure truly disappears and is no longer present. At first sight, one might be tempted to conclude that the structure is absent based on the stellar density distributions. However, upon closer inspection of the unsharp mask, there seems to be a faint peanut signal near the ends of the bar, leading us to lean towards the first option, that the structure weakens but is still present. A more detail study about this topic will be address in a future work.
%%%%%%%%%%%%%%%%%%%%%%%%%%%%%%%%%%%%%%%%%%%%%%%%%%

% Don't change these lines
\bsp	% typesetting comment
\label{lastpage}
\end{document}